\documentclass[%
 amsmath,amssymb,
 aps,
floatfix,
showkeys
]{revtex4-2}

\newcommand{\Sec}[1]{{Sec.~\ref{#1}}}
\newcommand{\Alg}[1]{{Algorithm~\ref{#1}}}

\usepackage{verbatim}
\usepackage{algorithm}
\usepackage{algpseudocode}
\usepackage{lineno}
\usepackage{placeins}
\usepackage[english]{babel}

\algnewcommand{\True}{\textbf{true}}
\algnewcommand{\False}{\textbf{false}}

\usepackage{graphicx}
\usepackage{dcolumn}
\usepackage{bm}

\begin{document}

\preprint{APS/123-QED}

\title{The influence of spatial configuration in collective transitions: \\the importance of being sorted}

\author{Daniel Galvis}

\affiliation{Centre for Systems Modelling and Quantitative Biomedicine, University of Birmingham, UK}
\affiliation{Institute of Metabolism and Systems Research (IMSR), University of Birmingham, UK}

\author{David J. Hodson}
\affiliation{Institute of Metabolism and Systems Research (IMSR), University of Birmingham, UK}
\affiliation{Centre of Membrane Proteins and Receptors (COMPARE), University of Birmingham, Birmingham, UK}
\affiliation{Centre for Endocrinology, Diabetes and Metabolism, Birmingham Health Partners, Birmingham, UK}
\affiliation{Oxford Centre for Diabetes, Endocrinology and Metabolism, Radcliffe Department of Medicine, University of Oxford, Oxford, OX3 7LE, UK}
\author{Kyle C.~A. Wedgwood}
\affiliation{Living Systems Institute, University of Exeter, UK}
\affiliation{EPSRC Hub for Quantitative Modelling in Healthcare, University of Exeter, UK}
\affiliation{College of Engineering, Mathematics and Physical Sciences, University of Exeter, UK}
\keywords{Excitable systems, collective dynamics, beta cell, sortedness}

\date{\today}

\begin{abstract}
We studied the effects of spatial configuration on collective dynamics in a nearest-neighbour and diffusively coupled lattice of heterogeneous nodes. The networks contained nodes from two populations, which differed in their intrinsic excitability. Initially, these populations were uniformly and randomly distributed throughout the lattice. We then developed an iterative algorithm for perturbing the arrangement of the network such that nodes from the same population were increasingly likely to be adjacent to one another. We found that the global input strength, or \textit{network drive}, necessary to transition the network from a state of quiescence to a state of synchronised and oscillatory activity was decreased as \textit{network sortedness} was increased. Moreover, for weak coupling, we found that regimes of partial synchronisation exist (i.e. 2:1 resonance in the activity of the two populations), which were dependent both on network drive (sometimes in a non-monotonic fashion) and network sortedness. 
\end{abstract}

\flushbottom
\maketitle
\thispagestyle{empty}

%
%

    

%
%

\section{Introduction}
\label{sec:intro}

Many nonlinear systems exhibit \textit{excitable} behaviour, whereby
they exhibit large-amplitude oscillations in response to
small-amplitude, transient perturbations. Such excitable dynamics are
observed in semiconductor lasers \cite{Terrien2020,Terrien2021}, social media networks \cite{Mathiesen2013}, epidemiology \cite{Vannucchi2004}, and
wildfires \cite{Punckt2015}. One prominent example is electrically excitable cells, such as neurons \cite{Izhikevich2000a,DeMaesschalck2015a,Wedgwood2021a}, cardiac cells \cite{Majumder2018,Barrio2020}, pituitary cells \cite{Sanchez-Cardenas2010,Hodson2012a} and
pancreatic beta cells \cite{Bertram2007a,McKenna2016a}. 
When excitable units are combined into networks, they can generate complex
rhythms \cite{Bittihn2017,Horning2017,Fretter2017}. Interestingly, such networks may also generate
dynamics that occur over low-dimensional manifolds of the full system \cite{Ashwin92,Watanabe1993,Ott2009,Bick2020}.
For example, neurons in the pre-B\"otzinger complex fire synchronously to induce the inspiratory
and expiratory phases during breathing \cite{Wittmeier2008,Gaiteri2011}.

Heterogeneity is ubiquitous in natural systems. 
Whilst often portrayed as a undesirable attribute, it can play an important role in governing network dynamics \cite{Manchanda2017,Delgado2018,Lambert2018}. 
For example, neurons may coarsely be stratified into excitatory and inhibitory groups, with the former promoting firing behaviour
in other neurons and the latter suppressing it. 
When coupled, these neuronal subtypes give rise to a variety of behaviours, including synchronisation, and enable the network to respond differentially to incoming inputs \cite{Borgers2003,Borgers2005,Kopell2010}. 
The classification of neuronal subtypes is becoming ever finer \cite{Gouwens2019,Lipovsek2021} and it remains an open question as to how this heterogeneity governs overall brain dynamics.
Even when networks comprise only a single unit type, heterogeneity may still impact the global dynamics.
For example, if the natural frequencies of nodes in a coupled oscillator network are too far apart, the network will be unable to synchronise and will instead display more complex rhythms \cite{Ottino-Loffler2016}.

Here, we explore transitions to synchrony in a locally-coupled
network of heterogeneous, excitable nodes. As a motivating example, we consider
networks of pancreatic beta cells. Individually,
these cells exhibit excitable dynamics akin to the
Hodgkin--Huxley model of nerve cells \cite{Hodgkin1952}. Cells remain at rest
until they receive a significantly large electrical impulse or the extracellular concentration of glucose surpasses a threshold value \cite{Ashcroft1989,Braun2008}. 
Under sustained suprathreshold stimulation, cells exhibit repetitive bursting-type dynamics comprising epochs of firing activity, followed by periods of rest \cite{Kinard1999b}.
Beta cells are arranged into diffusively, and locally-coupled networks via
channels known as gap junctions \cite{Rorsman2012,Benninger2011}.
These networks exhibit synchronous bursting activity when exposed to sufficiently high levels of glucose \cite{Markovic2015a}. 
Although exogenous factors, such as incretin \cite{Hodson2014a} and paracrine \cite{Caicedo2013} signalling influence this coordinated beta cell response, the importance of intercellular coupling has been highlighted in
several studies that demonstrate that synchronous beta cell rhythms are disrupted when gap junctions are
blocked \cite{Head2012,Benninger2014a}.

Based on empirical evidence from rodents, it has generally been assumed that beta cells form a \textit{syncticium}, such that the activity of the network can be described by a single cell \cite{Dolensek2013,Satin2020a,Podobnik2020}.  
Recent studies have challenged this perspective, highlighting that some `leader cells' disproportionately influence the activity of a entire network made up primarily of `follower cells' \cite{Johnston2016a,Westacott2017,Salem2019,Benninger2021}.
One hypothesis suggests that islets are composed of a small number ($\sim$10\%) of
highly excitable cells, with the remainder being less excitable \cite{Benninger2018b}. 
In this study, we explore how the spatial organisation of these two sub-populations affects the propensity of the whole network to oscillate in a synchronous fashion. 
The remainder of the manuscript is arranged as follows: In \Sec{sec:methods}, we describe the beta cell model, introduce a metric that captures how sorted a network is with respect to its heterogeneity, and present an algorithm that can generate networks with arbitrary sortedness.
In \Sec{sec:results}, we investigate how dynamic
transitions to synchronous bursting depends of the degree of sortedness in the network and end in \Sec{sec:discussion} with concluding remarks.

\section{Methods}
\label{sec:methods}

\subsection{Mathematical model}
We consider a network of $N$ diffusively-coupled excitable cells from a model describing electrical activity in pancreatic beta cells in the presence of glucose \citep{Sherman1988a}. These cells exhibit \textit{bursting} dynamics (in voltage $v$) when the glucose level, $G \in [0,1]$, is sufficiently high. The system possesses a slow variable, $c$, representing Ca\textsuperscript{2+} concentration, which oscillates when the cell is active. We arrange $N = 1,018$ nodes on a hexagonal close-packed (hcp) lattice embedded within a sphere. Each node is connected to its nearest neighbours via gap-junction coupling. The parameter $\overline{g}_L$ sets the excitability of single cells within the network (Fig. \ref{fig:single_cell}). We define two sub-populations of nodes distinguished by their excitability. Population 1 is highly excitable ($\overline{g}_L = 60$) and population 2 is less excitable ($\overline{g}_L = 100$). We then consider the range over $G$ where population 1 nodes are intrinsically active, while population 2 cells are intrinsically inactive. A full description of the mathematical model is provided in \Sec{sec:model}.

\subsection{Measuring sortedness}
\label{sec:assort}

To track the degree of sortedness in the network, we define a \textit{node sortedness} measure that, for a given node, measures the proportion of neighbours that are of the same population type. For a general network with nodes attributed to $K \in \mathbb{N}$ populations, the node sortedness, $A_i$, is defined as

\begin{equation}
    A_i = \frac{1}{|J_i|}\sum_{j\in J_i} \chi_{ij}, \quad \chi_{ij} = 
    \sum_{k=1}^K \mu_i^{(k)} \mu_j^{(k)}, \quad i = 1,\dots, N, \quad
    \mu_{i}^{(k)} =
    \begin{cases}
        1, & i \in P_k \\
        0, & \text{otherwise}
    \end{cases},
    \label{eq:node_assort}
\end{equation}

\noindent
where the \textit{population sets} $P_k$ contain the indices of the nodes within population $k = 1,\dots,K$ and form a partition over the node indices $\{ 1, 2, \dots, N\}$, $J_i$ is the set of indices of nodes that are adjacent to node $i$, $\mu_i^{(k)}$ is an indicator function that takes value 1 if $i$ belongs to population $k$ and value 0 otherwise,
and $\chi_{ij}$ is an indicator function that takes value 1 when node $i$ and $j$ belong to the same population and value 0 otherwise.
For each population, the average node sortedness is defined via

\begin{equation}
    \overline{A}_k = \frac{1}{|P_k|} \sum_{n \in P_k} A_n, \quad k = 1,2, \dots K.
    \label{eq:ave_node_assort}
\end{equation}

\noindent
Finally, the \textit{network sortedness} is defined as

\begin{equation}
    \mathcal{A} = \frac{1}{K-1}\left(-1 + \sum_{k=1}^{K}{\overline{A}_k}\right).
    \label{eq:net_assort}
\end{equation}

\noindent
where $\mathcal{A}\in [{-1}/(K - 1),1]$ and, for the present case with $K=2$, $\mathcal{A}\in [-1,1]$.
For a network in which populations are assigned to nodes following a uniformly random distribution,
 $\mathcal{A}\approx 0$ since $\overline{A}_{k}$ is approximately equal to ${N_k}/{N}$ where $N_k$, $k=1,2$ is the number of nodes in population $k$. 
 and therefore $\sum_{k}\overline{A}_{k} \approx 1$.
An illustration of the computation of the sortedness metrics \eqref{eq:node_assort}-\eqref{eq:net_assort} is shown in Fig.~\ref{fig:assortativity}.

\subsection{Modifying network sortedness}
\label{sec:algorithm}

Here, we describe our approach for generating networks with different network sortednesss.
The algorithm works by exchanging the population type of nodes from different populations randomly 
to increase (or decrease) $\mathcal{A}$.
The algorithm begins by randomly permuting the order of the $N$ indices.
The first $N_1$ indices of the permuted sequence are attributed to $P_1$, with the remaining $N_2$ indices attributed to $P_2$, yielding a distribution of population 1 nodes that is uniformly random in space.

On each iteration, $a$, of the algorithm, pairs of nodes (from different populations) are sampled without replacement from a joint probability density function (pdf) 

\begin{equation} 
    P(X=i, Y=j) = f(i,j), \quad i \in P_1, j \in P_2,
    \label{eq:pdf}
\end{equation}

\noindent
where $X$ and $Y$ are random integer variables indicating the node selected from population 1 and 2, respectively.
The population types of these nodes are then exchanged, that is, if $i \in P_1$ and $j \in P_2$, then $i$ is added to $P_2$ and removed from $P_1$ and vice versa for $j$.
The network sortedness \eqref{eq:net_assort} is then recomputed for the adjusted population sets.
If the exchange leads to an increase (decrease) in $\mathcal{A}$, the exchange is accepted and the algorithm proceeds to iteration $a+1$.
If the exchange does not lead to an increase (decrease) in $\mathcal{A}$, the exchange is rejected and indices $i$ and $j$ are placed back in $P_1$ and $P_2$, respectively.
In this case, a new pair of nodes is drawn from $f$ and the process is repeated until either: a pair whose exchange leads to an increase (decrease) in $\mathcal{A}$ is found and the algorithm proceeds to the next iteration; or it is determined that no such pair exists, at which point the algorithm terminates.
An example of one iteration of this algorithm is depicted in Fig.~\ref{fig:alg_iteration}.
We refer to the algorithm in which swaps are accepted only if they lead to an increase (decrease) in $\mathcal{A}$ as the \textit{forward} (\textit{backward}) algorithm. 
We define $\mathcal{A}_{a}$ to be the evaluation of $\mathcal{A}$ of the network after $a$ iterations.
Running the algorithm to convergence produces the sets $\mathcal{P}_{k} = \{P_{k}^{a}\}_{a=0}^{a_\text{final}}$ containing the population sets after each iteration. 

\subsubsection{Modified sortedness metrics}
\label{sec:modified_alg}

\begin{figure}[t] 
\centering
\includegraphics[width=0.35\linewidth]{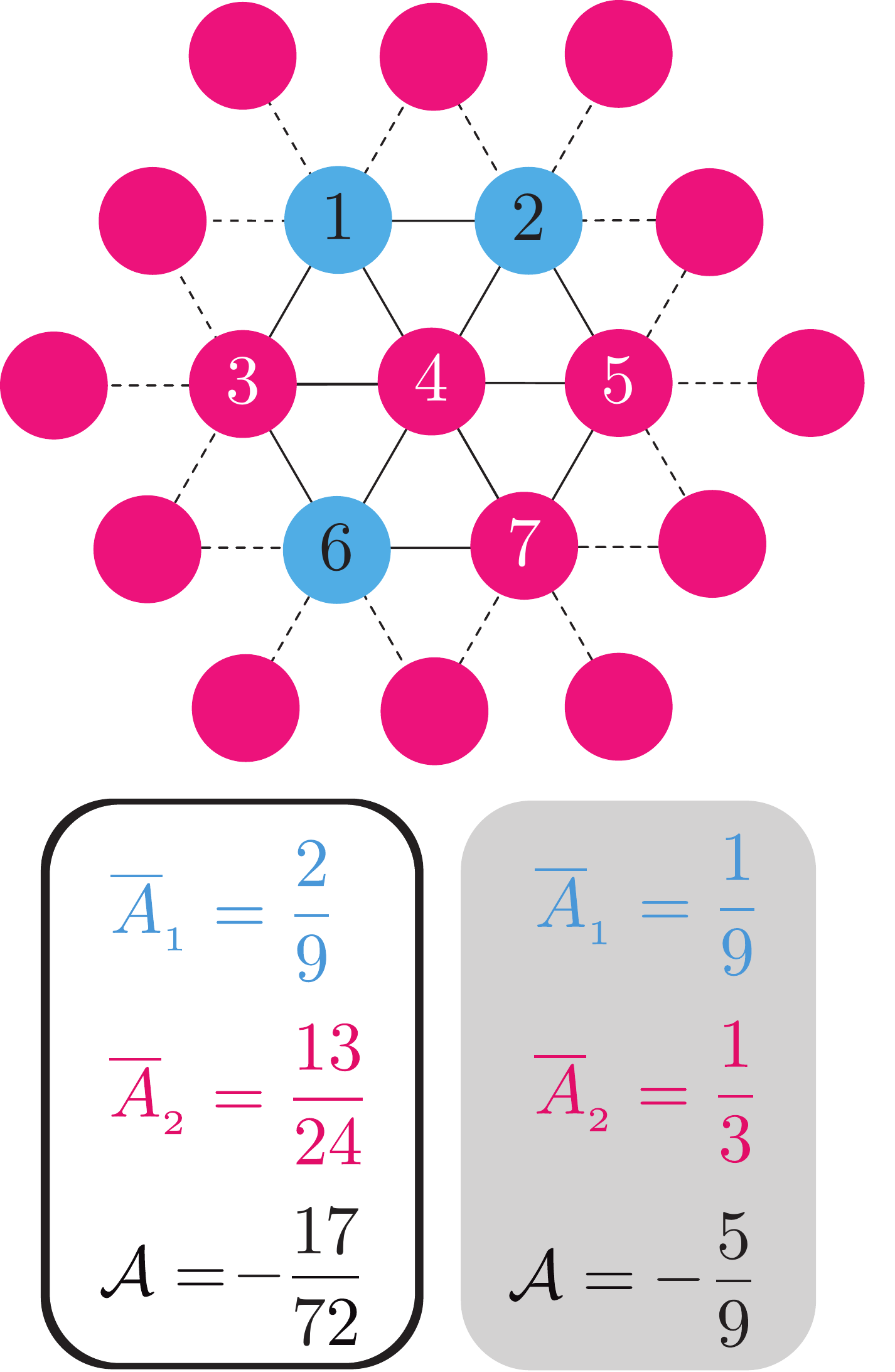}
\caption{\textbf{Example showing the modified sortedness metric.} The network under consideration is the interior portion of the depicted network with population sets $P_1 = \{1,2,6\}$ (blue) and $P_2 = \{3,4,5,7\}$ (pink). Using the original node sortedness metric \eqref{eq:node_assort}, the network sortedness as computed by \eqref{eq:net_assort} is $\mathcal{A}=-17/72$. The modified node sortedness \eqref{eq:node_assort_rev} assumes that each of the boundary nodes $i \in \{1,2,3,5,6,7\}$ has an additional $J-|J_i|$  connections to population 2 nodes, where $J_i$ is the set of nodes to which node $i$ was originally coupled. These additional connections are depicted by the dashed edges emanating from the boundary nodes. In this planar domain example, each of the boundary nodes has $|J_i|=3$ connections and $J=6$. Using the modified node sortedness metric, the network sortedness has value $\mathcal{A} = -5/9$. 
}
\label{fig:adjusted_assort}
\end{figure}

Although the algorithm yields well-sorted networks with a small number of clusters of nodes from population 1, these clusters preferentially form at the edges of the domain.
The average node sortedness, as defined in \eqref{eq:ave_node_assort}, for population 1 is maximised when a single cluster of population 1 nodes is coupled to the smallest possible number of population 2 nodes.
This naturally occurs at the edges of the domain, since any cluster of population 1 nodes in the domain interior must be surrounded by population 2 nodes.
We are interested in the dynamics that arise as the small population of highly excitable cells forms clusters within the lattice,
hence, we wish to remove this tendency for clusters to form at the domain boundary.
To overcome this, we use a modified definition of the node sortedness \eqref{eq:node_assort}

\begin{equation}
    \widetilde{A}_i = \frac{1}{J} \sum_{j\in J_i} \chi_{ij} + \frac{\mu_i^{(2)}\left(J - |J_i|\right)}{J}, \quad i = 1, \dots N,
    \label{eq:node_assort_rev}
\end{equation}

\noindent
where $J = 12$ is the number of connections that interior lattice nodes possess.
For nodes with $|J_{i}| < J$ (i.e., nodes on the domain boundary) the additional term in \eqref{eq:node_assort_rev} compared to \eqref{eq:node_assort} incorporates a further $J-|J_{i}|$ connections to population 2 nodes for the purposes of calculating node sortedness values.
This procedure is equivalent to assuming that the lattice defining our domain is embedded within a larger lattice of population 2 nodes.
An example of the computation of network sortedness using 
\eqref{eq:node_assort_rev} is shown in Fig.~\ref{fig:adjusted_assort}. Pseudocode for the network sortedness manipulation algorithm is provided in \Sec{sec:alg}.

\subsubsection{Node selection probabilities}
\label{sec:prob}

In this section, we formulate the node selection pdf used in the network sortedness adjustment algorithm.
We assume that the selection of node from $P_1$ is independent of the selection of node from $P_2$ so that \eqref{eq:pdf} becomes

\begin{equation}
    f(i,j) = f_{P_1}(i) f_{P_2}(j), \quad i \in P_1, j \in P_2.
\end{equation}

\noindent
One choice would set $f_1$ and $f_2$ to be uniform over $P_1$ and $P_2$, respectively.
Empirical observations of the algorithm outcome in this case demonstrate that clusters of population 1 nodes tend to form at the edge of the domain (not shown).
As discussed in \Sec{sec:modified_alg}, we wish to avoid this scenario.
The tendency for clusters to form near the edge occurs because of the spherical nature of our lattice domain.
In particular, a uniform choice for $f_1$ and $f_2$ means that nodes at the centre of the domain are less likely to be selected under a uniformly random sampling of indices than those at the edge because the number of nodes in the network increases superlinearly with respect to the domain radius.
Therefore, we derive choices for $f_{P_k}$ that equalise the probability of a node being selected on the basis of its radial coordinate. 
The heuristic for generating $f_{P_1}$ will be the same as that for generating $f_{P_2}$ up to the population identity.

Denote the radial distance from the origin of node $i\in \mathbb{N}_N$ by $r_i = (x_i^2 + y_i^2 + z_i^2)^{1/2} \in \mathbb{R}_{\geq 0}$ where $(x_i,y_i,z_i)\in \mathbb{R}^3$ are the Cartesian coordinates of the location of the node.
We define a sequences of intervals, $\mathcal{I}_n = [(n-1) \delta r, \, n\delta r]$, for $n=1,\dots 8$ with $\delta r = r_\text{max}/8$ where $r_\text{max} = \max_i\{ r_i\}$ so that each node is assigned to exactly one interval.
The set of nodes from $P_k$ belonging to a given interval $\mathcal{I}_n$ is given by $R_{n,P_k} = \{ i\in \mathbb{N}_N \mid r_i \in \mathcal{I}_n, \, i \in P_k\}$.
Using these set definitions, the pdf $f_{P_k}$ may be defined as

\begin{equation}
    f_{P_k}(i) = \frac{1}{Q|R_{n_i,P_k}|},
     \quad i \in \mathbb{N}_N.
     \label{eq:selection_weights}
\end{equation}

\noindent
where $R_{n_i,P_k}$ is such that $r_i \in \mathcal{I}_{n_i}$ and $Q$ is a normalisation factor ensuring that $\sum_{i \in P_k} f_{P_k}(i) = 1$.
This choice for $f_{P_k}$ reweights the probability of a given node being selected by a factor proportional to the number of cells from the same population within a spherical annulus with inner and outer radii specified by the boundaries of the intervals $\mathcal{I}_n$.
This reweighting favours selecting nodes closer to the centre of the domain over those more distal.

\subsection{Evaluation of collective dynamics}
\label{sec:sum_stats}

To characterise the network dynamics, we consider two features based on the  Ca\textsuperscript{2+} trajectories across all nodes, namely, the mean number of peaks ($\overline{P}$) and the time-averaged degree of synchronisation ($\overline{R}$) calculated as the average magnitude of the Kuramoto order parameter (\ref{eq:kuramoto}).
The mean number of Ca\textsuperscript{2+} peaks across all nodes is proportional to the \textit{network participation}, that is, the fraction of nodes that undergo oscillation.
The value of $\overline{R}$ captures the \textit{network coordination}, tracking how closely the phases of the Ca\textsuperscript{2+} trajectories stay to one another across the simulation duration. 
We additionally define $\overline{P}_{k}$ and $\overline{R}_{k}$ where $k \in \{1,2\}$ to be the mean number of peaks in Ca\textsuperscript{2+} and the time-averaged degree of synchronisation across nodes in population $P_k$, respectively.
\section{Results}
\label{sec:results}

\subsection{Simulating dynamics on the set of networks defined by running the swapping algorithm to convergence}
\label{subsec:single_network}
We ran the swapping algorithm to convergence (in the forward direction and with 10\% of the nodes specified to be from population 1) to produce sets $\mathcal{P}_{1}$ and $\mathcal{P}_{2}$.
For this run, the network configuration converged after $a_{final} = 203$ iterations with a corresponding sortedness of the terminal network configuration of $\mathcal{A}_{final} = 0.69$.
We then simulated the dynamical system \eqref{eq:Veq}-\eqref{eq:coupling} for $10$ equispaced values of $G \in [0.3,0.55]$, as described in \Sec{sec:sims} for $g_{coup} \in \{1,2,10\}$, and each configuration of populations defined by the population sets contained in $\mathcal{P}_{1}$ and $\mathcal{P}_{2}$ for $a \in \{1,4,7,\dots,a_{final}\}$.
We ran each simulation for $T_{max} = 360,000$ ms ($6$ minutes), and discarded the initial 90,000 ms of resulting times series to control for transients.
Each network configuration was simulated three times using each of a pre-defined set of initial conditions. 
Finally, we ran simulations once more using the first of these initial conditions 
to verify that results were consistent when the simulation duration was increased.
We then calculated the features $\overline{P}$ and $\overline{R}$ for each simulation.

To aid in interpreting the results, we define the following sets. Firstly, the parameter domain over which we evaluated the dynamical system was $\mathcal{D} = \{(\mathcal{A}, G) \mid \mathcal{A} \in[0,\mathcal{A}_{final}], G\in[0.3,0.55]\}$.
Secondly, the level sets $L^{+} = \{(\mathcal{A}, \ G) \mid \overline{P}(\mathcal{A}, \ G)=5\}$, $L^{*} = \{(\mathcal{A}, \ G) \mid \overline{R}(\mathcal{A}, \ G)=0.9\}$, and $L_k^{+} = \{(\mathcal{A}, \ G) \mid \overline{P_k}(\mathcal{A}, \ G)=5\}$ (for $k \in {1,2}$) were used to delineate subsets of $\mathcal{D}$ with qualitatively distinct network dynamics, which will be described below.


\subsubsection{Increasing $\mathcal{A}$ lowers the drive $G$ required for a transition to globally synchronised bursting when coupling is strong}
\label{subsubsec:strong_coupling}

\begin{figure}[ht] 
\centering
\includegraphics[width=0.5\textwidth]{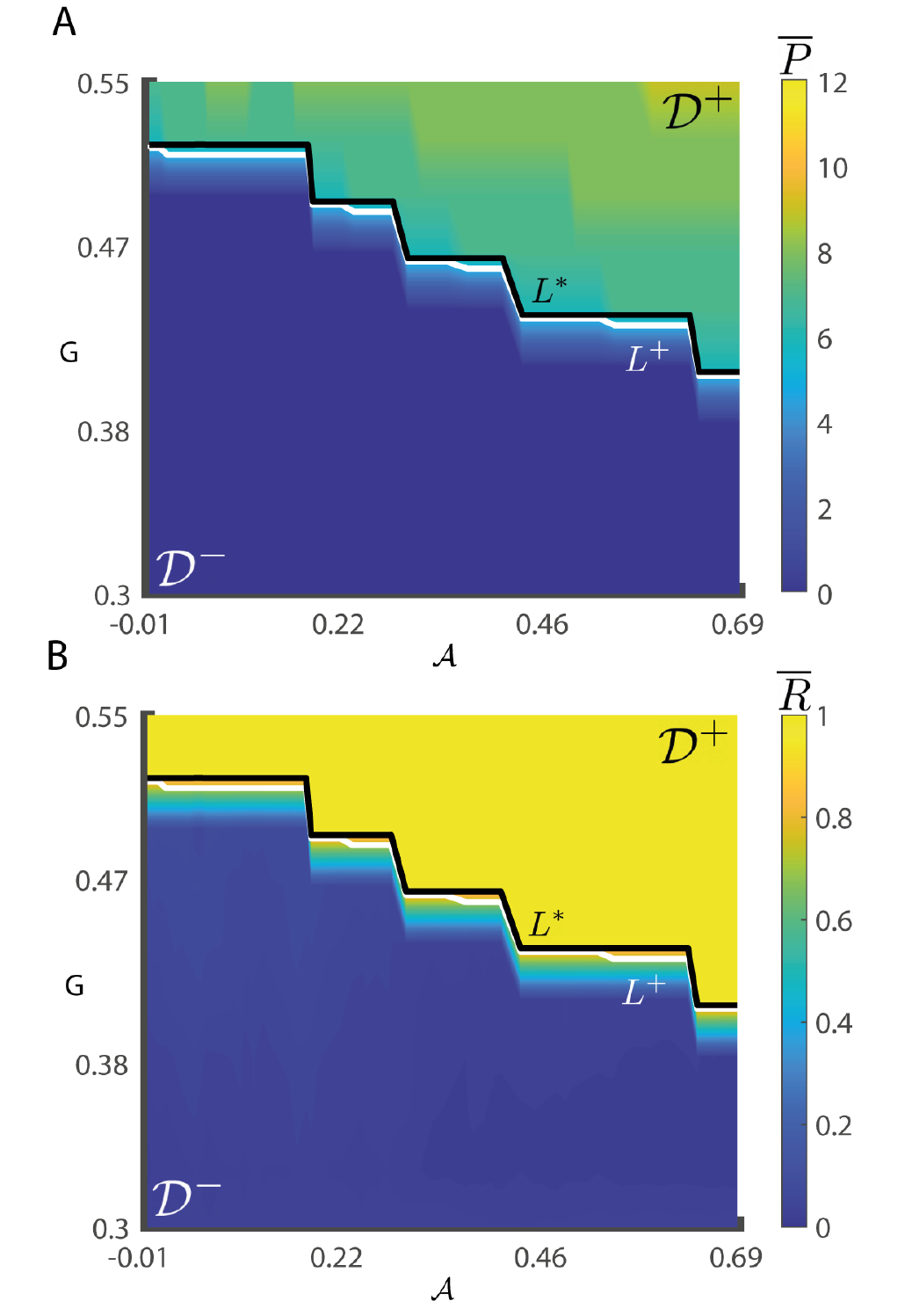}
\caption{\textbf{Network activity with respect to sortedness and drive for strong coupling.} \textbf{A)} Plotting $\overline{P}$ averaged over three sets of initial conditions shows that for increasing $\mathcal{A}$, lower drive $G$ is required to activate the network. \textbf{B)} Plotting $\overline{R}$ averaged over three sets of initial conditions shows that for increasing $\mathcal{A}$, lower drive $G$ is required to synchronise the network.}
\label{fig:network_set_strong_stripped}
\end{figure}

We first sought to establish whether there is a relationship between $\mathcal{A}$ and the level of drive $G$ required to activate the network.
Shown in Fig. \ref{fig:phase_transition_example} is an example in which the transition from global quiescence to global activation is dependent on both $G$ and $\mathcal{A}$ for the strongly coupled ($g_{coup} = 10$) case and where population 1 nodes comprise $10\%$ of the network.
The mean of the Ca$^{2+}$ trajectories for population 1 and 2 across the network are plotted for several values of $\mathcal{A}$ and $G$, which shows that as $\mathcal{A}$ increases, the required drive $G$ to activate the network decreases. 
To examine trends across a range of network configurations, we plot the features $\overline{P}$ (Fig.~\ref{fig:network_set_strong_stripped}A, Fig.~\ref{fig:network_set_strong}A), and $\overline{R}$ (Fig.~\ref{fig:network_set_strong_stripped}B, Fig.~\ref{fig:network_set_strong}B) as a function of both $G$ and $\mathcal{A}$. 
Each point depicts a value $S(\mathcal{A}, \ G)$, where $S \in \{\overline{P}, \overline{R}\}$, taken to be the median feature across the three simulations (which differ only in their initial condition).
For strong coupling, we found that $\mathcal{D}$ can be separated into a quiescent regime ($\mathcal{D}^{-}$) and an oscillatory one ($\mathcal{D}^{+}$).
The level set curve $L^{+}$ separating these regimes can be parameterised as a non-increasing function of $\mathcal{A}$ (i.e., $G = L^+(\mathcal{A})$), supporting the hypothesis that increasing $\mathcal{A}$ decreases the drive required for network activation (Fig.~\ref{fig:network_set_strong_stripped}A, B white curve).
Similarly, the level set $L^{*}$ can be parameterised as a non-increasing function of $\mathcal{A}$ that also separates the domains $\mathcal{D}^{-}$ and $\mathcal{D}^{+}$ (Fig.~\ref{fig:network_set_strong_stripped}A, B black curve).

To investigate the robustness of the above relationships, we plotted $\overline{P}$ and $\overline{R}$ resulting from each of the three initial conditions (Fig.~\ref{fig:network_set_strong_yi}). 
We defined
curves $L^{+}$ (Fig.~\ref{fig:network_set_strong_yi}, white curves) and $L^{*}$ (Fig.~\ref{fig:network_set_strong_yi}, black curves), in the same manner as described above.
These curves are not identical across the choices of initial condition, and both curves are non-monotonic for the third initial condition,  suggesting that multi-stability exists for some $(\mathcal{A}, \ G) \in \mathcal{D}$, at least near the transition between regimes $\mathcal{D}^{-}$ and $\mathcal{D}^{+}$.

\subsubsection{A domain with intra-population synchronicity and inter-population resonance exists when coupling is lowered to an intermediate strength}
\label{subsubsec:middle_coupling} 

\begin{figure}[ht] 
\centering
\includegraphics[width=0.48\textwidth]{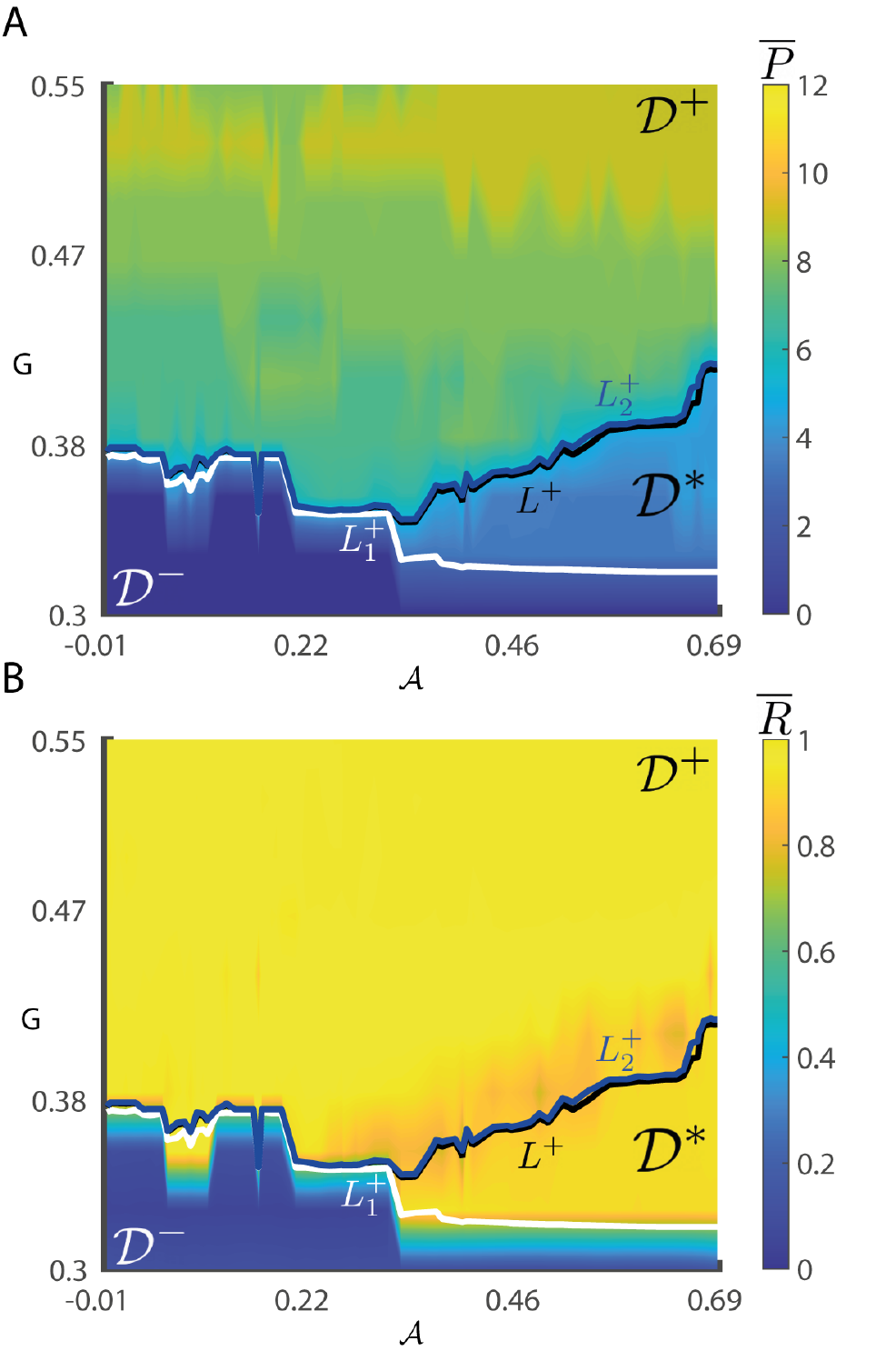}
\caption{\textbf{Network activity with respect to sortedness and drive for middle-strength coupling.} \textbf{A)} Plotting $\overline{P}$ averaged over three sets of initial conditions shows a third regime $\mathcal{D}^*$ bounded by $L_1^+$ and $L_2^+$. \textbf{B)} Plotting $\overline{R}$ averaged over three sets of initial conditions shows that for increasing $\mathcal{A}$, lower drive $G$ is required to synchronise the network.}
\label{fig:network_set_middle_stripped}
\end{figure}

When $g_{coup} = 2$ (intermediate strength coupling), the threshold for activation of the network was lower than in the case of strong coupling, owing to the reduction of the suppressing effect of the less excitable population 2 nodes on the more excitable population 1 nodes.
As in \Sec{subsubsec:strong_coupling}, we plot the features $\overline{P}$ (Fig.~\ref{fig:network_set_middle_stripped}A, Fig.~\ref{fig:network_set_middle}A) and $\overline{R}$ (Fig.~\ref{fig:network_set_middle_stripped}B, Fig.~\ref{fig:network_set_middle}B) within the parameter domain $\mathcal{D}$, taking the median across three initial conditions
We observed that the domain can be separated into three regimes with qualitatively distinct dynamics.
The first two regimes, $\mathcal{D}^{-}$ and $\mathcal{D}^{+}$, contain dynamics where the majority of nodes are quiescent or active (Fig.~\ref{fig:network_set_middle_stripped}A) and synchronised (Fig.~\ref{fig:network_set_middle_stripped}B), respectively.
Within the third regime, denoted $\mathcal{D}^{*}$, we found high intra-population synchronisation, with population 2 nodes oscillating (w.r.t. Ca\textsuperscript{2+}) at a frequency approximately half that of the population 1 nodes on average (Fig.~\ref{fig:network_set_middle}C, D triangle). 
i.e., this regime produces inter-population resonance at a 2:1 ratio.
Moreover, between the regimes $\mathcal{D}^*$ and $\mathcal{D}^+$, we found a sliver of the domain with lowered synchronisation (Fig.~\ref{fig:network_set_middle} star), where the population 2 oscillation frequency approaches that of population 1.
For low values of $\mathcal{A}$, the curves $L_{k}^{+}$ nearly overlap one another and separate the regimes $\mathcal{D}^{-}$ and $\mathcal{D}^{+}$, however, for larger values of $\mathcal{A}$, these curves diverge and bound the $\mathcal{D}^{*}$ regime. 
Due to the large fraction of nodes being contained in population population 2, we find that the curve $L^{+}$, defined as in \Sec{subsubsec:strong_coupling}, approximately overlaps $L_{2}^{+}$.

The curve $L_{1}^{+}$ marks the transition from quiescence to activity, which may or may not be synchronised, and is non-monotonic. 
Despite this non-monotonicity, there still exists an overall trend linking increases in $\mathcal{A}$ and the required drive to induce activity, $G$.
In particular, for larger values of $\mathcal{A}$, where increasing $G$ results in a transition to $\mathcal{D}^{*}$, the required drive to pass through $L_{1}^{+}$ is lowest. 
Moreover, when $\mathcal{A}$ is near $\mathcal{A}_{0}$, i.e., at early iterations of the algorithm, the required drive to pass through $L_{1}^{+}$ is highest (Fig.~\ref{fig:network_set_middle_stripped}A, B). 

When redefining the curves $L_{k}^{+}$ for $k \in \{1,2\}$ and $L^{+}$ for individual sets of initial conditions, we again found that they were not identical, implying the presence multi-stability near the transitions between regimes. 
In addition, we also found cases of multi-stability within the regime $\mathcal{D}^{+}$ (Fig.~\ref{fig:network_set_middle_yi}G, H).
For example, Fig.~\ref{fig:network_set_middle_yi} shows the plots of $\overline{P}$ (Fig.~\ref{fig:network_set_middle_yi}A, C, E) and $\overline{R}$ (Fig.~\ref{fig:network_set_middle_yi}B, D, F) resulting from each initial condition separately.
For some points $(\mathcal{A},G)$, we observed lower synchronisation ($\overline{R}$) for some initial conditions (Fig.~\ref{fig:network_set_middle_yi} square) relative to the others (Fig.~\ref{fig:network_set_middle_yi} circle). 
These points of lowered synchrony persisted when $T_{max}$ was increased suggesting that this activity was not due extended transient behaviour (not shown).

\subsubsection{When coupling strength is low, only population 1 activation depends on sortedness}
\label{subsubsec:low_coupling} 

\begin{figure}[ht] 
\centering
\includegraphics[width=0.55\textwidth]{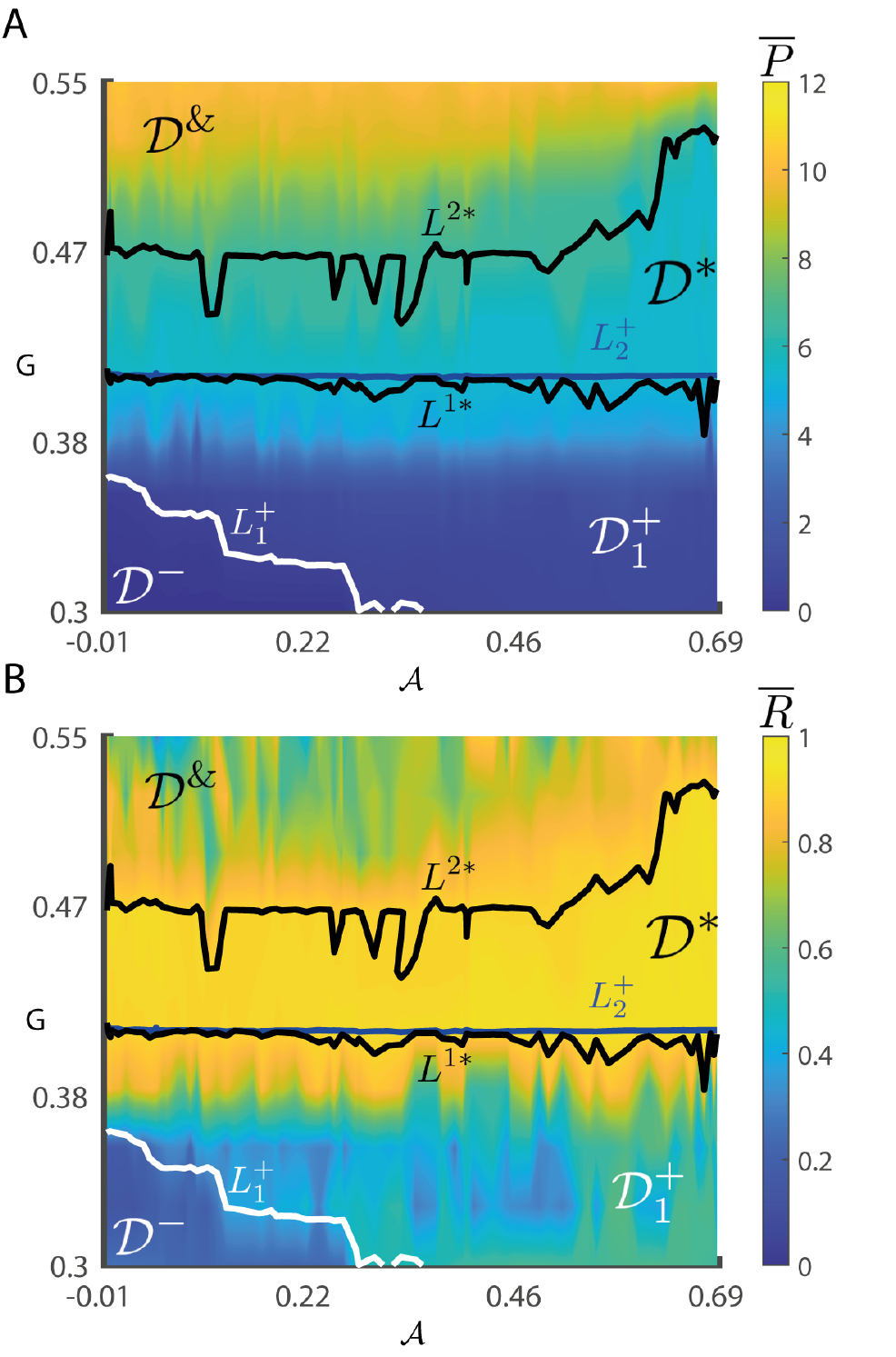}
\caption{\textbf{Network activity with respect to sortedness and drive for weak coupling.} \textbf{A)} Plotting $\overline{P}$ averaged over three sets of initial conditions shows that activation of population 1, but not population 2, is dependent on sortedness. \textbf{B)} Plotting $\overline{R}$ averaged over three sets of initial conditions shows that synchronisation is non-monotonic with respect to $G$, peaking within a $2:1$ resonance regime $\mathcal{D}^*$.}
\label{fig:network_set_weak_stripped}
\end{figure}

When $g_{coup} = 1$ (low coupling strength), we observed a greater variety of of parameter regimes supporting distinct dynamics (Fig.~\ref{fig:network_set_weak_stripped}, Fig.~\ref{fig:network_set_weak}) than in either the intermediate strength or strong coupling cases.
For this coupling strength, there is no regime in which the network is active and synchronised (i.e., regime $\mathcal{D}^{+}$ does not exist).
The region $\mathcal{D}^{-}$, in which the majority of nodes are inactive, exists for low values of $\mathcal{A}$ and $G$, and is bounded above by the curve $L_{1}^{+}$.

We next identified the regime $\mathcal{D}_{1}^{+}$ in which only nodes in $P_{1}$ are active while those in $P_{2}$ remain silent (Fig.~\ref{fig:network_set_weak}, circle).
In this regime, population 1 nodes are active but only weakly coordinated while population 2 nodes are mostly inactive (Fig.~\ref{fig:network_set_weak}D).
This results in a weak global signal (low amplitude oscillations of the average Ca\textsuperscript{2+} signal) (Fig.~\ref{fig:network_set_weak}C).
This regime can bounded below by $L_{1}^{+}$ and above by $L_{2}^{+}$ and also by $L^{+}$ (not shown).
A region, denoted $\mathcal{D}^{*}$, also exists with similar dynamics to the one defined for $g_{coup} = 2$. 
Within this regime, the overall network synchronisation is high (Fig.~\ref{fig:network_set_weak_stripped}B), however, population 2 nodes exhibit oscillatory Ca$^{2+}$ behaviour with approximately half the frequency of that of the population 1 nodes (Fig.~\ref{fig:network_set_weak}C, D triangle). 

A final region, $\mathcal{D}^{\&}$, exists for high values of $G$, where network synchronisation decreases (Fig.~\ref{fig:network_set_weak_stripped}B) whilst the average number of peaks continues to increase (Fig.~\ref{fig:network_set_weak_stripped}A). 
The level set $L^{*}$ defines two separate curves, labelled $L^{1*}$ and $L^{2*}$, due to the non-monotonicity of the synchronisation index with respect to $G$.
The curve $L^{2*}$ bounds $\mathcal{D}^{*}$ from above and separates it from $\mathcal{D}^{\&}$, whilst $L^{1*}$ is a lower bound for $\mathcal{D}^{*}$ and separates it from $\mathcal{D}_{1}^{+}$.
Fig.~\ref{fig:network_set_weak}C (star) shows the irregular global signal caused by weak coordination, which is shown in Fig.~\ref{fig:network_set_weak}D (star).
As in the case for $g_{coup} = 2$, the curve $L_{1}^{+}$ marks the transition from quiescence to activity, however, in this case only population 1 nodes become active. 
This curve is non-monotonic, however, the overall trend once again links increases in $\mathcal{A}$ with a lower required drive to induce activity. 
On the other hand, the curve $L_{2}^{+}$ does not appear to be dependent on $\mathcal{A}$.
Additionally, we found that the curve $G = L^{2*}(\mathcal{A})$, defined by the set $L^{2*}$, shows an increasing trend, which suggests that the range of $G$ for which maximal synchronisation occurs increases with $\mathcal{A}$.

\section{Discussion}
\label{sec:discussion}

In this manuscript, we demonstrated how transitions to globally-coordinated activity are dependent on the degree of sortedness in  population excitability.
We used a prototypical model of a pancreatic beta cell where a small population was highly excitable, whilst a larger population was less excitable.
As the global drive to the network was increased, activity across the network transitioned from a globally inactive state to one in which subsets of nodes became active and synchronised their activity.
By perturbing the spatial distribution of the highly excitable population, we showed that the drive strength at which such transitions occur is dependent on the sortedness of the network.
These results have specific implications for insulin secretion in the pancreatic islets of Langerhans, and more general implications regarding transitions to synchrony and other forms of collective dynamics in networks of coupled excitable units.

To perform our study, we developed \Alg{alg:orig}, which perturbs the sortedness of the network in a directed manner.
Whilst our algorithm is tailored towards spherical geometries and local, diffusive coupling, it can be adapted to other geometries and coupling types, since the neighbourhoods can be succinctly encoded in the adjacency matrix.
In addition, although our study focused on conditions in which there are only two different populations, \Sec{sec:methods} discusses how our metrics can be extended to networks with more population types.
Given the growing interest in studying heterogeneous populations in complex networks, we hope that our algorithms will prove useful to other researchers in the future.

\section{Acknowledgements}
\label{sec:acknowledgements}

DG acknowledges funding from the University of Birmingham Dynamic Investment Fund and the EPSRC Centre grant EP/N014391/2.
DJH acknowledges funding from the MRC Projects MR/N00275X/1 and MR/S025618/1 the and Diabetes UK Project Grants 17/0005681. 
This project has also received funding from the European Research Council (ERC) under the European Union’s Horizon 2020 research and innovation programme (Starting Grant 715884 to DJH).
KCAW acknowledges funding from the MRC Fellowship MR/P01478X/1 and the Hub for Quantitative Modelling in Healthcare EP/T017856/1.

\bibliography{IsletHeterogeneity}

\newcommand{\beginsupplement}{%
        \setcounter{table}{0}
        \renewcommand{\thetable}{S\arabic{table}}%
        \setcounter{figure}{0}
        \renewcommand{\thefigure}{S\arabic{figure}}%
        \setcounter{section}{0}
        \renewcommand{\thesection}{S\arabic{section}}%
        \setcounter{equation}{0}
        \renewcommand{\theequation}{S\arabic{equation}}%
     }
     
\beginsupplement
\section{Supplemental material}
\label{sec:supplemental}

\subsection{Mathematical model}
\label{sec:model}

We consider a network of $N$ diffusively coupled excitable cells, each of which is described by the three variable model

\begin{align}
C_m \frac{dV_i}{dt}  &= - I_{K}(V_i,n_i) - I_{Ca}(V_i,h_i) - I_{K-Ca}(V_i,c_i) - I_{L}(V_i) - I_{coup,i}, \quad i = 1,\dots N,\label{eq:Veq} \\
\frac{dn_i}{dt} & = \frac{n_{\infty}(V_i) - n_i}{\tau_{n}(V_i)}, \label{eq:neq} \\
\frac{dc_i}{dt} & = - f (\alpha I_{Ca}(V_i,c_i) + k_{Ca} c_i). \label{eq:ceq}
\end{align}

\noindent
This system was adapted from the Sherman--Rinzel--Keizer model, which describes the dynamics of electrical activity in pancreatic beta cells in the presence of glucose \citep{Sherman1988a}. The intrinsic dynamics of the voltage, $V$ given by \eqref{eq:Veq} are driven by K\textsuperscript{+} ($I_{K}$), Ca\textsuperscript{2+} ($I_{Ca}$), and Ca\textsuperscript{2+}-activated K\textsuperscript{+} ($I_{K-Ca}$) ionic currents, with a rate governed by the whole cell capacitance given by $C_m$.
These currents are described via

\begin{align}
    I_K(V,n) &=  \overline{g}_{K} n (V - V_{k}), \label{eq:IK} \\
    I_{Ca}(V,h) &= \overline{g}_{Ca} m_{\infty}(V)h_{\infty}(V)(V - V_{Ca}), \\
    I_{K-Ca}(V,c) &= \overline{g}_{K-Ca} \frac{c}{K_{d} + c} (V-V_{k}), \\
    I_{L}(V) &= \overline{g}_{L} (1-G) (V - V_{K}) \label{eq:leak}.
\end{align}

\noindent
In \eqref{eq:IK}-\eqref{eq:leak}, $\overline{g}_X$ denotes the maximal conductance of the channel $X$ where $X \in \{ K, Ca, K-Ca, L\}$ where $L$ signifies a leak channel; $V_X$ are the reversal potentials of the respective channels, $m$ and $n$ are the proportion of open activating gates for the Ca$^{2+}$ and K$^{+}$ channels, respectively; $h$ is the proportion of open inactivating Ca$^{2+}$ channels; $c$ is the cytosolic concentration of Ca$^{2+}$; and $G$ is the extracellular concentration of glucose, which provides a global drive to promote activity and is taken to be homogeneous across the network. 
The activation of $I_{K-Ca}$ is a function of free intracellular Ca\textsuperscript{2+} concentration and is defined by a Hill-type function with disassociation constant $K_d$.
The current $I_{coup,i}$ captures the influence of the coupling between cells and will be discussed in \Sec{sec:coupling}.

The dynamics for $n$ and $h$ follow exponential decay to their state values given by

\begin{equation}
x_{\infty}(V)  = \frac{1}{1 + \exp{[(V_{x} - V) / S_{x}]}}, \quad x \in \{h,m,n\}, \label{eq:ststgate}
\end{equation}

\noindent
at a rate given by the voltage-dependent time constant

\begin{equation} \label{eq:taun}
\tau_{n}(V) = \frac{\overline{\tau}}{\exp{[(V-\overline{V}) / \kappa_1]} + \exp{[-(V-\overline{V}) / \kappa_2]}} .
\end{equation}

\noindent
In \eqref{eq:ststgate}, $V_x$ represents the activation (inactivation) thresholds for $m$ and $n$ ($h$) and $S_x$ represents the sensitivity of the channels around this point. Finally, \eqref{eq:ceq} describes the evolution of the concentration of cytosolic Ca$^{2+}$, which decays and is pumped out of the cell following a combined linear process with rate $k_{Ca}$ and enters the cell via the Ca$^{2+}$ ion channel at a rate given by the scale factor $\alpha$. The parameter $f$ specifies the fraction of free to bound Ca$^{2+}$ in the cell, where the bound Ca$^{2+}$ plays no role in the relevant dynamics in our model.

The electrical activity of pancreatic beta cells is proportional to the extracellular concentration of glucose.
For sufficiently high extracellular glucose, the cells exhibit \textit{bursting} dynamics, in which their voltage periodically switches between high frequency oscillations and quiescence.
The high frequency oscillations in voltage are correlated with the secretion of insulin from these cells, so that these bursting dynamics are tightly coupled to the cells' functional role.
To expose the dependence of our system on glucose, we introduced a hyperpolarising leak current given by \eqref{eq:leak} that explicitly depends on the glucose concentration $G$.
For an isolated cell (i.e., without coupling) with the parameters specified in Table~\ref{tbl:parameters}, the system describing each node exhibits steady state behaviour for low $G$ and passes through a bifurcation as $G \in [0,1]$ is increased, as shown in Fig.~\ref{fig:single_cell}.

The bursting dynamics in our model are of the \textit{fold-homoclinic} type under the classification specified in \cite{Izhikevich2000a}.
This classification is based on separation of the full system into a fast subsystem \eqref{eq:Veq}-\eqref{eq:neq} and a slow subsystem \eqref{eq:ceq}, treating the slow subsystem variables (in this case, $c$) as parameters in the fast subsystem.
During each bursting cycle, the slow evolution of $c$ pushes the fast subsystem through bifurcations that initiate and terminate oscillatory behaviour.
In particular, when $c$ decreases to a small enough value, the fast subsystem passes through a fold bifurcation in which a stable steady state and a saddle steady state collide and annihilate one another.
Following this,the system exhibits stable periodic activity, during which $c$ increases according to \eqref{eq:ceq}.
When $c$ increases to a sufficiently large value, the fast subsystem passes through a homoclinic bifurcation that destroys the periodic orbit and the system returns to the original stable steady state. 
Following this, $c$ decreases until it once again reaches the fold point and the cycle repeats.

\begin{table}
\centering

 \begin{tabular}{|c c | c c | c c |} 

 \hline
 Parameter & Value & Parameter & Value & Parameter & Value \\ [0.5ex] 
 \hline
 $C_{m}$ (fF) & 5310 & 
 $V_{m}$ (mV) & 4  &
 $S_{m}$ (mV) & 14 \\
 $V_{n}$ (mV) & -15 &
 $S_{n}$ (mV) & 5.6 &
 $\kappa_1$ (mV) & 65 \\
 $\kappa_2$ (mV) & 20 &
 $\overline{\tau}$ (ms) & 37.5 &
 $\overline{V}$ (mV) & -75 \\
 $V_{h}$ (mV) & -10 &
 $S_{h}$ (mV) & -10 &
 $\overline{g}_{K}$ (pS) & 2500 \\
 $\overline{g}_{Ca}$ (pS) & 1400 &
 $V_{K}$ (mV) & -75 &
 $V_{Ca}$ (mV) & 110 \\
 $K_{d}$ ($\mu$ M) & 100 &
 $\overline{g}_{K-Ca}$ (pS) & 30000 &
 $f$ & 0.001\\
 $k_{Ca}$ (ms$^{-1}$) & 0.03 &
 $\alpha$ $\left(\frac{\mu \text{m}^{3} \text{Coul}}{\text{mMol}}\right)$ & $4.5061\times 10^{-6}$  &
 $\overline{g}_{coup}$ (pS) & \{varies\} \\
\hline

\end{tabular}
 \caption{Parameter values of the oscillator model.}
 \label{tbl:parameters}
\end{table}

\subsection{Model simulations}
\label{sec:sims}

Simulations were conducted using Matlab 2019B. The dynamical systems were solved using ode15s, the relative tolerance set to $10^{-5}$, and explicit Jacobians were provided. The code was run on the University of Birmingham BlueBEAR HPC running RedHat 8.3 (x86\_64)(see http://www.birmingham.ac.uk/bear for more details). Each set of simulations ran over 16 cores using a maximum of 128GB RAM (32GB was sufficient in most cases). All code used in the project is freely available for download from: github.com/dgalvis/network\_spatial.
\subsubsection{Initial Conditions}
Initial conditions $y_i(0) = \left(V_{i}(0), n_{i}(0), c_{i}(0)\right)$ for node $i=1,\dots,N$ were sampled independently from the distributions
\begin{equation}
    V_i(0) \sim \mathcal{N}(-68, (68/6)^2), \quad n_i(0) = 0, \quad  c_{i}(0) \sim \mathcal{N}(0.57, (0.57/6)^2),
\end{equation}

\noindent
where $\mathcal{N}(\mu,\sigma^2)$ represents a normal distribution with mean $\mu$ and variance $\sigma^2$.
Throughout, we use $Y(0)$ to denote the set of initial conditions across the whole network, i.e., $Y(0) = \left( y_1(0), \dots, y_N(0)\right)$.

\subsubsection{Excitability and drive in the single-cell model}

\begin{figure}[t!] 
\centering
\includegraphics[width=1.0\textwidth]{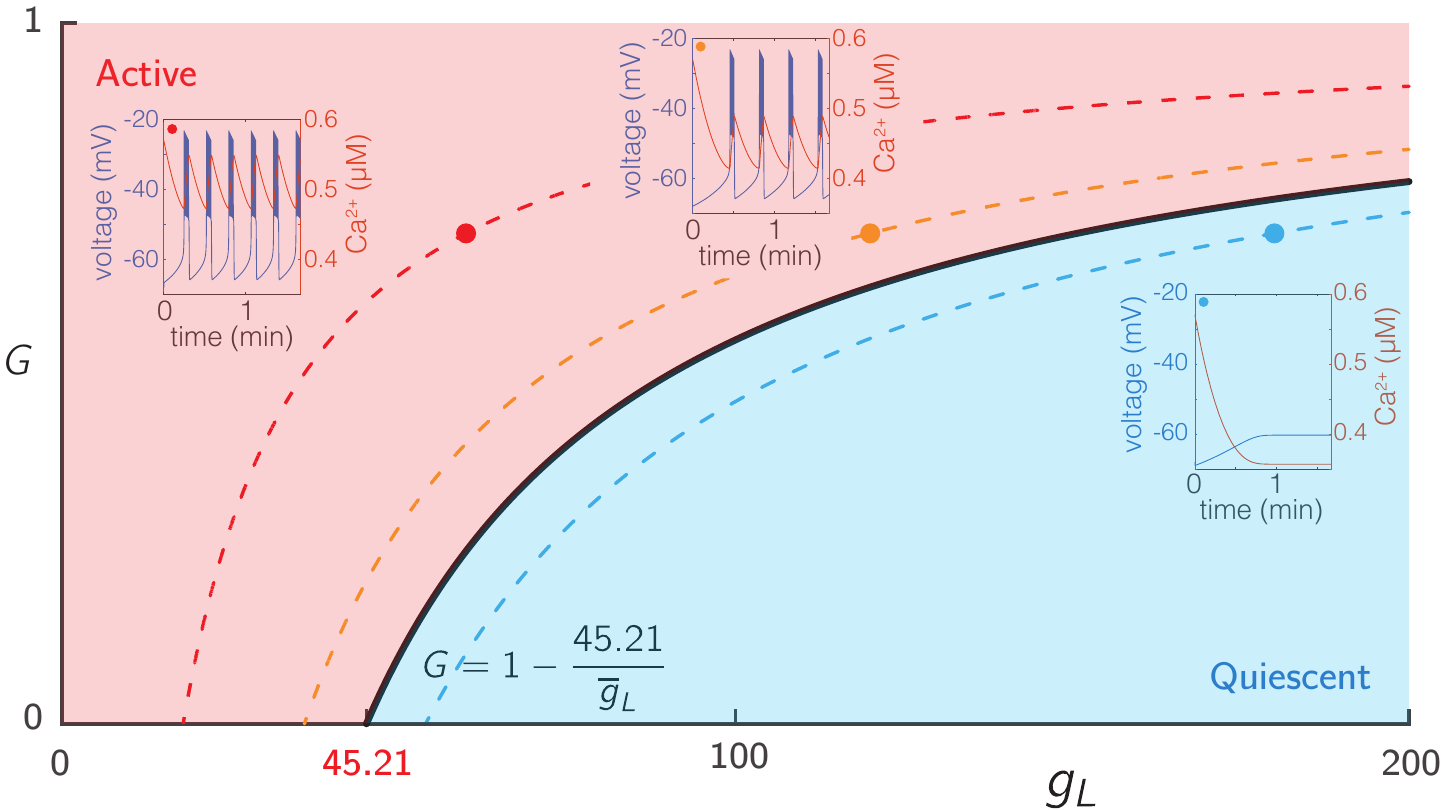}
\caption{\textbf{Excitability of single cells.} The voltage traces from three cells with varying levels of intrinsic excitability ($\overline{g}_{L}$), but the same level of drive ($G = 0.7$). The red, blue, and black traces show decreasing levels of excitability with values of $\overline{g}_{L} = 60$, $120$, and $180$, respectively. More excitable cells have a shorter interburst interval. The solid black line represents a Hopf bifurcation as a function of both $G$ and $g_{L}$. At the lowest drive ($G = 0$), the Hopf bifurcation occurs for $\overline{g}_{L} = 45.21$ pS.  The dotted lines represent ``level sets" of the $(\overline{g}_L,G)$ parameter space, along which the excitability of the single cell is identical. Data for the bifurcation diagram was computed using XPP 8.0 \citep{Ermentrout2002}.}
\label{fig:single_cell}
\end{figure}

The ionic current $I_L$ (\ref{eq:leak}) is a hyperpolarising current that can be used to adjust the excitability of each cell and to determine the activation level of the network.
In particular, the maximum  conductance $\overline{g}_{L}$ determines the excitability of a cell. As this value increases, the cell becomes less excitable, that is, for a given value of $G$, cells with higher $\overline{g}_L$ are less likely to burst.
This behaviour is summarised in Fig.~\ref{fig:single_cell}, which shows a two parameter bifurcation diagram showing the transition from quiescent to bursting behaviour under simultaneous variation of $(\overline{g}_L,G)$, which occurs via a Hopf bifurcation of the full system \eqref{eq:Veq}-\eqref{eq:ceq}.
For $G=0$, this Hopf bifurcation occurs at $\overline{g}_{L} = 45.21$ pS.
For non-zero values of $G$, the bifurcation curve is defined via $(1 - G)\overline{g}_{L} = 45.21$ pS, as can be seen by examining the form of the \eqref{eq:Veq} and \eqref{eq:leak}.
Note that when $G = 1$, system \eqref{eq:Veq}-\eqref{eq:ceq} matches that of \cite{Sherman1988a}.
In the network modelling approach, we use the observations about the link between $\overline{g}_L$ and excitability to partition the network into two sub-populations, one being highly excitable, the other being significantly less excitable.

\subsubsection{Network structure and coupling}
\label{sec:coupling}
Pancreatic beta cells are arranged into roughly spherical clusters called islets of Langerhans (which also encompass other cell types which are disregarded in our model), which each contain $\sim$ 1,000 beta cells.
To capture this, we arrange $N = 1,018$ nodes on a hexagonal close packed (hcp) lattice embedded within a sphere.
The dominant form of coupling between beta cells in the islets is through gap junctions, which allow small molecules, including charged ions to pass directly from a cell to its adjacent neighbours.
Mathematically, this is represented through the inclusion of the diffusive term $I_{coup,i}$ in \eqref{eq:Veq} that factors in the local nature of coupling
\begin{equation}
I_{coup,i} = \overline{g}_{coup} \sum_{j\in {J}_i} (V_i-V_j), \label{eq:coupling}
\end{equation}

\noindent
where ${J}_i$ is the set of all cells to which cell $i$ is coupled.
Each node is connected to all of its nearest-neighbours so that the number of connections of nodes away from the boundary of the sphere is equal to the coordination number 12 whilst nodes on the boundary have fewer connections.

\subsubsection{Heterogeneity}
\label{sec:hetero}

We consider networks consisting of two sub-populations of nodes distinguished by their excitability (i.e., by their $\overline{g}_L$ values).
Population 1 is highly excitable ($\overline{g}_{L} = 60$ pS) and population 2 is less excitable ($\overline{g}_{L} = 100$ pS).
We then consider the range over $G$ for which population 1 nodes are intrinsically active (i.e., when $\overline{g}_{coup} = 0$) and population 2 nodes are intrinsically quiescent.
We then consider the effects of population size (by varying the proportion of overall network that population 1 nodes account for), the degree of sortedness between the two subpopulations (see \Sec{sec:assort}), global network drive ($G$), and global coupling strength on the collective dynamics of the network.

\newpage
\subsection{Description of the routines used by \Alg{alg:orig}}
\label{sec:alg}
\begin{algorithm}[H]
    \caption{\label{alg:orig} Algorithm for producing networks}
    \begin{algorithmic}[1]
    \scriptsize
    \Require
    \Statex $N$: number of nodes in network
    \Statex $a$: number of iterations of swapping algorithm to attempt
    \Statex $Dir$: signed integer determining whether algorithm runs forwards (positive) or backwards (negative)
    \Statex $\rho$: proportion of population 1 nodes
    \Ensure
    \Statex $\mathcal{A}$: network sortedness value
    \Statex $P_1$: population 1 set
    \Statex $P_2$: population 2 set
    \Statex $n$: number of swaps performed
    \Statex
    \Function{GenerateNetwork}{$N$, $a$, $Dir$, $\rho$}
         \State $(x,y,z)$, $r$, $K \gets$ \Call{EstablishLattice}{$N$}
         \State $N_1$, $N_2$, $P_1$, $P_2$, $\mathcal{A} \gets$ \Call{AssignInitialPopulations}{$N$, $\rho$}
         \State $Term \gets $ \False \Comment{Boolean determining whether terminal network state has been reached}
         \State $n \gets 0$
         \While{($n < a$) and ($Term =$ \False)}
             \State $f$, $F$, $Q \gets$ \Call{ComputeSelectionProbabilities}{$r[]$, $N_1$, $N_2$, $P_1$, $P_2$}
             \State $m \gets 0$
             \State $swap \gets$ \True \Comment{Boolean determining whether to attempt swaps}
             \While{($m < N_1 \times N_2$) and ($swap =$ \True)}
                 \State $\widetilde{P}_1$, $\widetilde{P}_2$, $\mathcal{A}_p$, $k \gets$\Call{NodeSwap}{$f$, $F$, $Q$, $Dir$, $N_1$, $N_2$, $P_1$, $P_2$}
                 \If{$\text{sgn}(\mathcal{A}_p - \mathcal{A}) = \text{sgn}(Dir)$}
                     \State $P_1$, $P_2 \gets \widetilde{P}_1$,  $\widetilde{P}_2$
                     \State $\mathcal{A} \gets \mathcal{A}_p$
                     \State $n \gets n+1$
                     \State $swap \gets$ \False
                 \Else  \Comment{Reject swap if $\mathcal{A}$ does not change in the desired direction}
                    \For{$l \gets k$ to $N_1 \times N_2$}
                        \State $F[l] \gets F[l]-f[k]$
                    \EndFor
                    \State $Q \gets Q - f[k]$
                    \State $m \gets m+1$
                \EndIf
            \EndWhile
            \If{$swap = $ \True}
                \State $Term \gets$ \True \Comment{Terminal state has been reached}
            \EndIf
        \EndWhile
        \State \Return $\mathcal{A}$, $P_1$, $P_2$, $n$
    \EndFunction
    \end{algorithmic}
\end{algorithm}

\Alg{alg:hcp_lattice1}-\Alg{alg:swap_nodes} are used by \Alg{alg:orig} which is described in the main text.

\Alg{alg:hcp_lattice1} returns a set of points in $\mathbb{R}^3$ corresponding to the centres of spheres within a hexagonal close packed lattice (hcp). The input $r_{ball}$ corresponds to the radius of the spheres within the lattice, which we set to $r_{ball}=0.5$ so that the distance between any two nearest neighbors is $d_{ball}=2r_{ball}=1$. \Alg{alg:hcp_lattice1} produces the hcp lattice using a sequence of scalings and shifts of a square lattice which takes the points $\{(x,y,z) \mid x,y,z \in \{1,\dots,M\}\}$, where $M$ is an integer corresponding to the number of spheres along the length of the lattice. We sought to embed a larger sphere, $S_{net}$, of radius $R_{net}$ within the resulting hcp-lattice, and therefore, must choose $M$ such that $S_{net}$ is contained within the lattice. For the square lattice, a natural choice would be $M=2R_{net}$, so that the length of the lattice equals the diameter of the sphere. However, for the hcp-lattice, the size of the resulting structure is $(M-1)x_{scale}+d_{ball} = Mx_{scale} = Md_{ball}$ by $(M-1)y_{scale}+d_{ball} > My_{scale}$  by $(M-1)z_{scale}+d_{ball} > Mz_{scale}$  (ignoring the shifts). To counteract this, we use:
\begin{equation}
    M = \text{ceil}\left(\frac{2R_{net}}{ \text{min}([x_{scale},y_{scale},z_{scale}])}\right).
\end{equation}
We found that this choice of $M$ generated a lattice which could fully embed the sphere, at least for our selection of $R_{net} = 5.55$ (in particular, we increased $M$ and found that the number of nodes within the sphere did not increase).

\Alg{alg:sphere_lattice} first runs \Alg{alg:hcp_lattice1} to produce an hcp-lattice. It then centres the lattice at the origin (i.e., at $(0,0,0)$) and finds all points that are within a sphere of radius $R_{net}$ centred at the origin, which define the nodes in the network. It also returns $N$, the number of nodes in the spherical hcp-lattice ($N = 1,018$ in this work). \Alg{alg:init_lattice} establishes the Boolean adjacency matrix representing the connections between nodes in the spherical hcp-lattice. A connection exists between two nodes if they are at a distance of $d_{ball}$ from one another. In other words, if two spheres (of radius $r_{ball}$) centred at the locations assigned to two nodes would be touching, then a connection exists between them. \Alg{alg:init_pop} determines the population sets $P_k$ for $k \in {1,2}$. It returns the number of nodes $N_k$ in each population, the population membership sets, and the initial network sortedness value $A_0$. \Alg{alg:build_f} determines the selection probabilities for every pair of nodes ($\{(i,j) \mid i \in P_1$, $j \in P_2\}$). \Alg{alg:swap_nodes} chooses a candidate swap, produces the population sets established by that swap, and calculates $\mathcal{A}$ for the updated population sets. 
 
\newpage
\begin{algorithm}[H]
    \caption{\label{alg:hcp_lattice1} Initialising HCP lattice}
    \begin{algorithmic}[1]
    \scriptsize
    \Require
    \Statex $R_{net}$: radius of the spherical lattice
    \Statex $r_{ball}$: radius of balls around points in the lattice
    \Ensure
    \Statex $(x_1,y_1,z_1), \dots, (x_N,y_N,z_N)$: $(x,y,z)$ coordinates of nodes 
    \Statex $N$: number of nodes in hcp-lattice
    \Statex
    \Function{EstablishHCPLattice}{$R_{net}, r_{ball}$}
    
    \State $d_{ball} \gets 2r_{ball}$
    \State $x_{scale} \gets d_{ball}$
    \State $y_{scale} \gets \sqrt{d_{ball}^2 - r_{ball}^2}$
    \State $z_{scale} \gets \sqrt{\frac{2}{3}}d_{ball}$
    \State $x_{shift} \gets r_{ball}$
    \State $y_{shift} \gets - \frac{d_{ball}}{\sqrt{3}}$

    \State $M \gets \text{ceil}(\frac{2R_{net}}{ \text{min}([x_{scale},y_{scale},z_{scale}])})$ 
    \State $counter \gets 0$
    
    \For{$i \gets 1$ to $M$}  
        \For{$j \gets 1$ to $M$}  
            \For{$k \gets 1$ to $M$} 
                \State $counter \gets counter + 1$
                \State $x[counter] \gets k \times x_{scale}$
                \State $y[counter] \gets j \times y_{scale}$
                \State $z[counter] \gets i \times z_{scale}$
                \If{$j$ even}
                    \State $x[counter] \gets x[counter] + x_{shift}$
                \EndIf
                \If{$i$ even}
                    \State $y[counter] \gets y[counter] + y_{shift}$
                \EndIf
            \EndFor
        \EndFor
    \EndFor
    \State $N \gets M^3$ \Comment{Total number of nodes in the lattice}
    \State \Return $(x,y,z)$, $N$
    \EndFunction
    \end{algorithmic}
\end{algorithm}

\begin{algorithm}[H]
    \caption{\label{alg:sphere_lattice} Initialising Sphere lattice}
    \begin{algorithmic}[1]
    \scriptsize
    \Require
    \Statex $R_{net}$: radius of the spherical lattice
    \Statex $r_{ball}$: radius of points in the lattice
    \Ensure
    \Statex $(x_1,y_1,z_1), \dots, (x_{N_{net}},y_{N_{net}},z_{N_{net}})$: $(x_{sphere},y_{sphere},z_{sphere})$ coordinates of nodes 
    \Statex $r_1, \dots, r_{N_{net}}$: $r_{sphere}$ radii of nodes
    \Statex $N_{net}$ number of nodes in the spherical lattice
    \Statex
    \Function{EstablishSphereLattice}{$R_{net}, r_{ball}$}
    \State $(x, y, z), N \gets $\Call{EstablishHCPLattice}{$R_{net}, r_{ball}$}
    \State $x \gets x - \text{mean}(x)$ \Comment{demean vector x}
    \State $y \gets y - \text{mean}(y)$ \Comment{demean vector y}
    \State $z \gets z - \text{mean}(z)$ \Comment{demean vector z}
    \State $r \gets \sqrt{x^2 + y^2 + z^2}$ \Comment{compute norm over all points}
    \State $counter \gets 0$
    \For{$i \gets 1$ to $N$}
        \If{$r[i] <= R_{net}$} \Comment{Find members of hcp-lattice within sphere radius $R_{net}$}
            \State $counter \gets counter + 1$
            \State $x_{sphere}[counter] \gets x[i]$
            \State $y_{sphere}[counter] \gets y[i]$
            \State $z_{sphere}[counter] \gets z[i]$
            \State $r_{sphere}[counter] \gets r[i]$
        \EndIf
    \EndFor
    \State $N_{net} \gets counter$ \Comment{Define number of nodes within the spherical domain}
    \State \Return $(x_{sphere},y_{sphere},z_{sphere})$, $r_{sphere}$, $N_{net}$
    \EndFunction
    \end{algorithmic}
\end{algorithm}

\begin{algorithm}[H]
    \caption{\label{alg:init_lattice} Initialising lattice}
    \begin{algorithmic}[1]
    \scriptsize
    \Require
    \Statex $R_{net}$: radius of the spherical lattice
    \Statex $r_{ball}$: radius of points in the lattice
    \Ensure
    \Statex $(x_1,y_1,z_1), \dots, (x_N,y_N,z_N)$: $(x,y,z)$ coordinates of nodes
    \Statex $r_1, \dots, r_N$: $r$ radial coordinate of nodes
    \Statex $K \in \mathbb{R}^N \times \mathbb{R}^N$: connectivity matrix
    \Statex
    \Function{EstablishLattice}{$R_{net}, r_{ball}$}
    \State $(x, y, z), r, N \gets $\Call{EstablishSphereLattice}{$R_{net}, r_{ball}$}
    \State $d_{ball} \gets 2 r_{ball}$
    \For{$i \gets 1$ to $N$}
        \For{$j \gets 1$ to $N$}
            \State $dist \gets \sqrt{(x[i] - x[j])^2 + (y[i] - y[j])^2 + (z[i] - z[j])^2}$
            \If{$dist = d_{ball}$} 
                \State $K[i][j] \gets 1$
            \Else
                \State $K[i][j] \gets 0$
            \EndIf
        \EndFor
    \EndFor
    \State \Return $(x,y,z)$, $r$, $K$
    \EndFunction
    \end{algorithmic}
\end{algorithm}

\begin{algorithm}[H]
    \caption{\label{alg:init_pop}  Initialising populations}
    \begin{algorithmic}[1]
    \scriptsize
    \Require
    \Statex $N$: number of nodes in network
    \Statex $\rho$: proportion of population 1 nodes
    \Ensure
    \Statex $N_1$, $N_2$: number of nodes in the respective population 1
    \Statex $P_1$, $P_2$: population sets
    \Statex $\mathcal{A}$: network sortedness
    \Statex
    \Function{AssignInitialPopulations}{$N$, $\rho$}
    \State $U \gets \text{random permutation of } \{1,\dots, N\}$
    \State $N_1 \gets \text{floor}(\rho N)$
    \State $N_2 \gets N - N_1$
    \State $P_{1}, P_{2} \gets$ integer array of length  $N_1$, integer array of length $N_2$
    \For{$k \gets 1$ to $N_1$}  
        \State $P_1[k] = U[k]$ \Comment{Assign first $N_1$ elements of $U$ to $P_1$}
    \EndFor
    \For{$k \gets 1$ to $N_2$} 
        \State $P_2[k] = U[N_1+k]$ \Comment{Assign last $N_2$ elements of $U$ to $P_2$}
    \EndFor
    \State $\mathcal{A} \gets $ network sortedness value \eqref{eq:net_assort} using $P_1$ and $P_2$
    \State \Return $N_1$, $N_2$, $P_1$, $P_2$, $\mathcal{A}$
    \EndFunction
    \end{algorithmic}
\end{algorithm}

\begin{algorithm}[H]
    \caption{\label{alg:build_f} Defining node pair selection probabilities}
    \begin{algorithmic}[1]
    \scriptsize
    \Require
    \Statex $r_1,\dots,r_N$: radial coordinates of nodes
    \Statex $N_1$, $N_2$: number of nodes in the respective population
    \Statex $P_1$, $P_2$: population sets
    \Ensure
    \Statex $f \propto$ probability density function for node pair selection
    \Statex $F \propto$ cumulative density function for node pair selection
    \Statex $Q$: normalisation constant for $f$
    \Statex
    \Function{ComputeSelectionProbabilities}{$r[]$, $N_1$, $N_2$, $P_1$, $P_2$}
    \State $f \gets $ array of length $N_1\times N_2$,
    \State $F \gets$  array of length $N_1\times N_2 +1$
        \State $F[1] \gets 0$
        \State $k$, $Q \gets 0$
        \For{$i \gets 1$ to $N_1$}
            \For{$j \gets 1$ to $N_2$}
            \State $k \gets k+1$
            \State $p \gets 1/R_{n_i,P_1} \times 1/R_{n_j,P_2}$ \Comment{Weight probability of node pair being selected using \eqref{eq:selection_weights}}
            \State $f[k] = p$
            \State $Q \gets Q + p$
            \State $F[k] \gets Q$
            \EndFor
        \EndFor
    \State \Return $f$, $F$, $Q$
    \EndFunction
    \end{algorithmic}
\end{algorithm}

\begin{algorithm}[H]
    \caption{\label{alg:swap_nodes} Node population swapping}
    \begin{algorithmic}[1]
    \scriptsize
    \Require
    \Statex $f \propto$ probability density function for node pair selection
    \Statex $F \propto$ cumulative density function for node pair selection
    \Statex $Q$: normalisation constant for $f$
    \Statex $P_1$, $P_2$: sets of indices of nodes in the respective population
    \Ensure
    \Statex $\widetilde{P}_1$, $\widetilde{P}_2$: population sets following node population swap
    \Statex $\mathcal{A}_p$: network sortedness of network with node populations swapped
    \Statex $k$: index of node pair swapped
    \Statex
    \Function{NodeSwap}{$f$, $F$, $Q$, $N_1$, $N_2$, $P_1$, $P_2$}
        \State $u \gets U(0,1)$ \Comment{Sample from unit uniform distribution}
        \State $k \gets 1$
        \While{$u < F(k)/Q$}
            \State $k \gets k + 1$
        \EndWhile
        \State $i$, $j \gets k/N_2$,  $(k-1)\!\!\! \mod N_2 + 1$ \Comment{Indices of selected population nodes}
        \State $\widetilde{P}_1$, $\widetilde{P}_2 \gets P_1$, $P_2$ \Comment{Create copies of $P_1$ and $P_2$} 
        \State $\widetilde{P}_1(i)$, $\widetilde{P}_2(j) \gets P_2(j)$, $P_1(i)$ \Comment{Trial node population swap}
        \State $\mathcal{A}_{p} \gets$ network sortedness value \eqref{eq:net_assort} using $\widetilde{P}_1$ and $\widetilde{P}_2$
        \State \Return $\widetilde{P}_1$, $\widetilde{P}_2$, $\mathcal{A}_p$, $k$
    \EndFunction
    \end{algorithmic}
\end{algorithm}

\subsection{Evaluation of collective dynamics}
\label{sec:sum_stats_supp}
For each node, the number of peaks was identified by searching for maxima exceeding 0.01 $\mu M$ in the Ca\textsuperscript{2+} timecourse across the simulation duration (see Fig. \ref{fig:single_cell}).

For a network with $N$ nodes, the time-dependent Kuramoto order parameter is a complex-valued scalar defined as 
\begin{equation}
{z}(t) = R(t) \mathrm{e}^{i\Theta(t)} = \frac{1}{N}\sum_{j=1}^{N}{\mathrm{e}^{i\theta_{j}(t)}}, 
    \label{eq:kuramoto}
\end{equation}
where $\theta_j(t)$ is the phase of the $j$th node, as extracted via a mean-subtracted Hilbert transform of the Ca\textsuperscript{2+} signal for node $j$.
The argument of $z$, $\Theta$, is the mean phase of the network whilst its magnitude, $R$, measures the degree of synchrony across the network.
We sample the Ca\textsuperscript{2+} at equispaced time points $t_i = i \delta t$, $i=0,\dots T-1$ and record the time-averaged degree of synchronisation: $\overline{R} = \frac{1}{T} \sum_{i=0}^{T-1} R(t_i)$.

\subsection{The swapping algorithm generally converges to a single cluster of population 1 nodes}

\begin{figure}[ht] 
\centering
\includegraphics[width=1.0\linewidth]{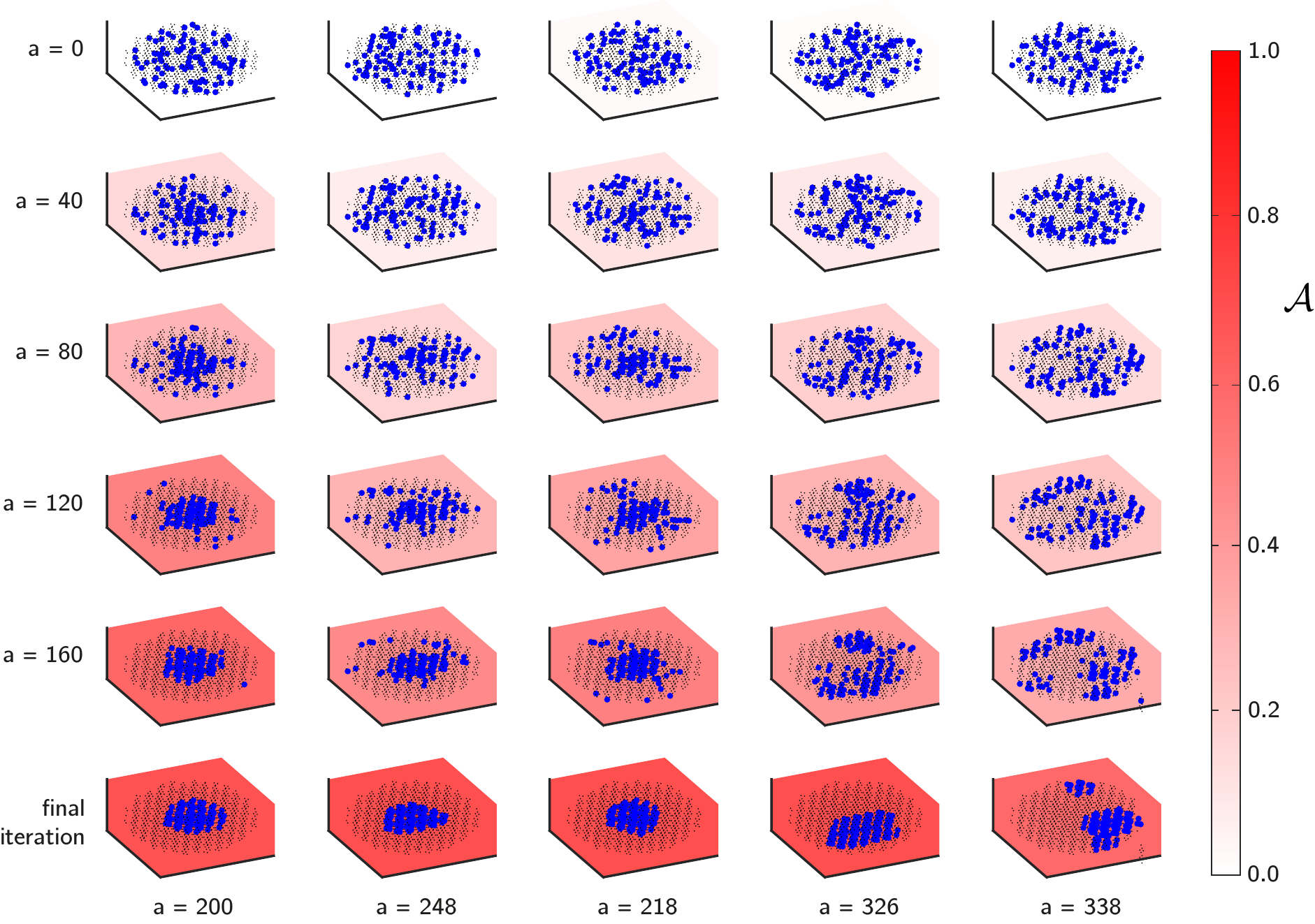}
\caption{\textbf{Examples of the swapping algorithm.} Five examples of the forward swapping algorithm and the associated $\mathcal{A}$ values. The final row of panels shows the final iteration, when no increases in $\mathcal{A}$ are possible. Population 1 nodes are $10\%$ of the total number of nodes and are shown in blue. Population 2 is shown in black. Generally, population 1 forms a single cluster as the algorithm converges, however this is not always the case (see example 5).}
\label{fig:assort_examples}
\end{figure}

\begin{figure}[ht] 
\centering
\includegraphics[width=1.0\linewidth]{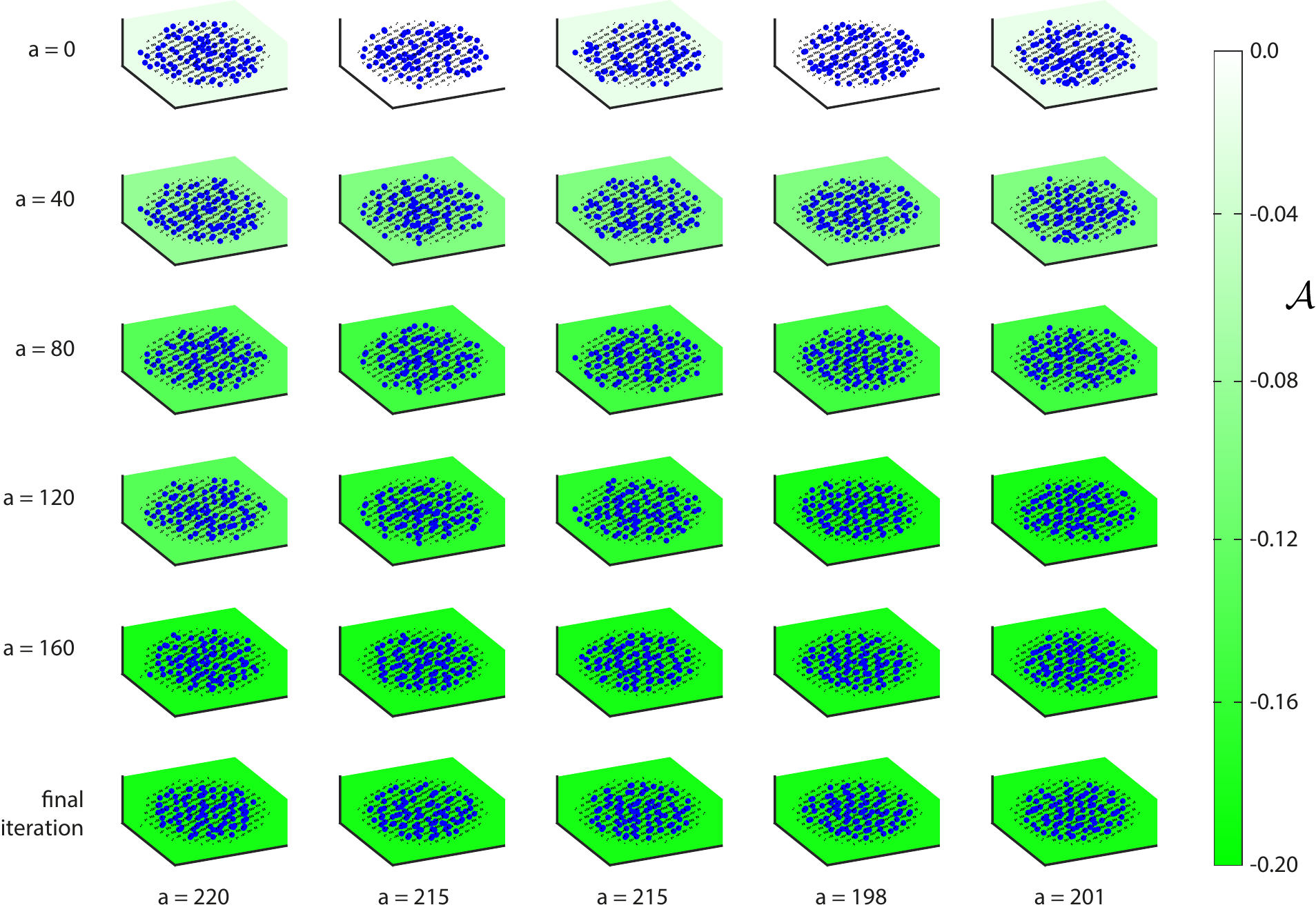}
\caption{\textbf{Examples of the backward swapping algorithm.} Five examples of the backward swapping algorithm and the associated $\mathcal{A}$ values. The final row of panels shows the final iteration, when no decreases in $\mathcal{A}$ are possible. Population 1 nodes account for $10\%$ of the total number of nodes and are shown in blue. Population 2 is shown in black. Generally, all population 1 nodes become isolated as the algorithm converges.}
\label{fig:assort_examples_backward}
\end{figure}

\begin{figure}[ht] 
\centering
\includegraphics[width=1.0\linewidth]{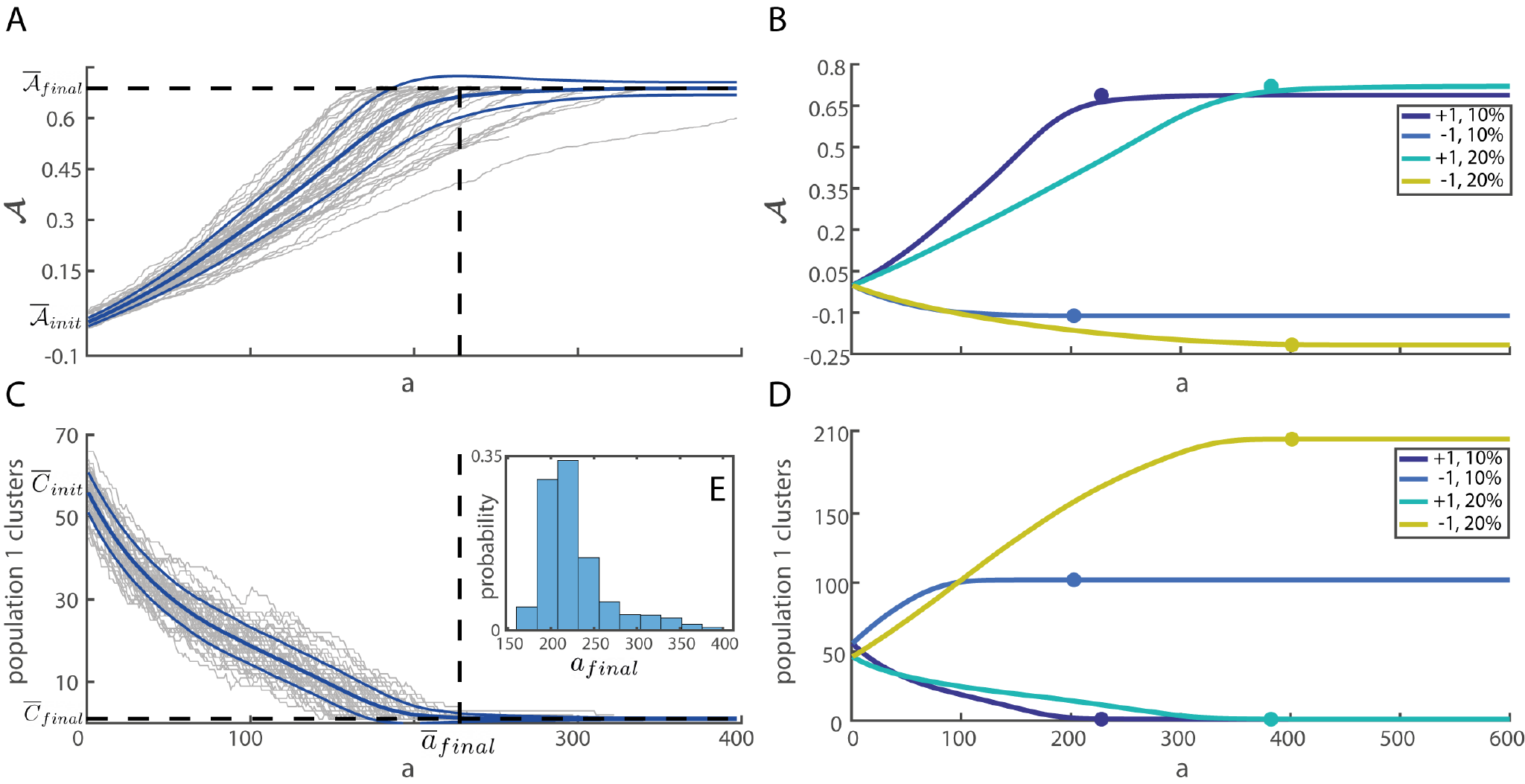}
\caption{\textbf{Convergence of the swapping algorithm.} \textbf{A)} The relationship between $\mathcal{A}$ and $a$ is shown for a subset of the 1,000 runs of \Alg{alg:orig} (grey lines). 
Population 1 was 10\% of the total number of nodes in the network. 
The average $\pm$ standard deviation is shown as blue lines.
$\overline{\mathcal{A}}_{init} = -9.37E-4$ was the mean value $\mathcal{A}$ when $a = 0$ over all runs.
$\overline{\mathcal{A}}_{final} = 0.69$ was the mean value of $\mathcal{A}$ when $a = a_{final}$ over all runs.
$\overline{a}_{final} = 227.75$ was the mean value of $a$ at $a_{final}$ over all runs.  
\textbf{B)} The relationship between population 1 clusters and $a$ is shown for a subset of the 1,000 runs of \Alg{alg:orig} (grey lines).  
Population 1 was 10\% of the total number of nodes in the network. The average $\pm$ standard deviation is shown as blue lines.
$\overline{C}_{init} = 56.02$ was the average number of population 1 clusters at $a = 0$ over all runs.
$\overline{C}_{final} = 1.05$ was the average number of population 1 clusters at $a = a_{final}$ over all runs.
\textbf{C)} The relationship between $\mathcal{A}$ and $a$ for the forward and backward algorithm when population 1 was $10\%$ and $20\%$ of the overall network.
The average over all runs is plotted for each case. 
Each curve also has a point of the same colour which indicates ($\overline{a}_{final}$, $\overline{\mathcal{A}}_{final}$). 
\textbf{D)} The relationship between population 1 clusters and $a$ for the forward and backward algorithm when population 1 was $10\%$ and $20\%$ of the overall network.
The average over all runs is plotted for each case. 
Each curve also has a point of the same colour which indicates ($\overline{a}_{final}$, $\overline{A}_{final}$).
\textbf{E)} An inset showing the distribution of $a_{final}$ over the 1,000 runs when population 1 was $10\%$ of the network and the algorithm was run in the forward direction.
}
\label{fig:assort_examples2}
\end{figure}

We first ran \Alg{alg:orig} $1,000$ times in configurations where nodes from population 1 accounted for $10\%$ of the network (i.e., $N_{1} = 102$ and $N_{2} = 916$).
The initial networks ($a = 0$) were uniform-randomly distributed ($\mathcal{A}_{init} = -9.37E-4 \pm 0.012$), and the algorithm was run until it reached convergence ($a=a_{final}$). Fig. \ref{fig:assort_examples} shows five examples (one per column) at several iterations between uniformly random spatial distribution ($a = 0$) and convergence ($a = a_{final}$). We found that convergence took $227.75 \pm 40.34$ iterations (Fig. \ref{fig:assort_examples2}E) and the final network sortedness was $\mathcal{A}_{final} = 0.69 \pm 0.019$. Fig. \ref{fig:assort_examples2}A shows examples of the relationship between $\mathcal{A}$ and $a$ for individual runs of the swapping algorithm (grey lines) as well as the average $\pm$ standard deviation (blue lines) over all the runs.

For each run, we determined the number of population 1 clusters (or connected components) as a function of iterations $a$. 
We found that the population 1 nodes were initially separated into $56.02 \pm 4.86$ connected components ($C_{init}$) at $a = 0$.
In $96.3\%$ of cases, population 1 formed a single cluster at $a = a_{final}$.
The first four columns in Fig. \ref{fig:assort_examples} show cases where population 1 converged to a single cluster. 
In the remaining $3.7\%$ of cases, the population 1 nodes formed multiple clusters at convergence.
One such examples of this is displayed in the fifth column in Fig. \ref{fig:assort_examples}, in which the final network consisted of three clusters. 
Across all runs, we found that the number of population 1 clusters at convergence was two, three, and four in $2.4\%$, $1.2\%$, and $0.1\%$ of runs, respectively.
Fig. \ref{fig:assort_examples2}C shows examples of the relationship between number of population 1 clusters and $a$ (grey lines) as well as the average $\pm$ standard deviation (blue lines) over all the runs.

We next ran the backward algorithm $1,000$ times when population 1 formed $10\%$ of the network nodes. Fig. \ref{fig:assort_examples_backward} shows five examples (one per column) at several iterations between uniform-random spatial distribution ($a = 0$) and convergence ($a = a_{final}$).
We found that convergence took $202.68 \pm 14.58$ iterations and the final network sortedness was $\mathcal{A}_{final} = -0.11 \pm 0.00$.
In addition, the number of connected components at $a_{final}$ was $102$ in each case. 
This is because the algorithm always reached a state in which all population 1 cells were isolated from one another (i.e., these nodes were coupled only to nodes from population 2). 
Finally, we ran the forward and backward algorithm again $1,000$ times when population 1 comprised $20\%$ of the network (i.e., $N_{1} = 204$ and $N_{2} = 814$).
The statistics for each of these cases are reported in Table \ref{tbl:stats}.
In Fig. \ref{fig:assort_examples2}B, we show the average relationship between $\mathcal{A}$ and $a$ and Fig. \ref{fig:assort_examples2}D shows the average relationship between population 1 clusters and $a$ over all runs for each case.

\begin{table}
\centering

\begin{tabular}{||c c c c c c c||} 
 \hline
 $N_1$ & $N_2$ & direction & $a_{final}$ & $\mathcal{A}_{final}$ & $C_{init}$ & $C_{final}$\\ [0.5ex] 
 \hline\hline
 102 & 916 & $+1$ & $227.75 \pm 40.34$ & $0.69 \pm 0.019$  & $56.02 \pm 4.86$ & $1.05 \pm 0.28$\\ 
 \hline
 102 & 916 & $-1$ & $202.68 \pm 14.58$ & $-0.11 \pm 0.00$ & $56.02 \pm 4.86$ & $102 \pm 0$ \\ 
 \hline
 204 & 814 & $+1$ & $382.36 \pm 56.12$ & $0.72 \pm 0.0060$  & $46.33 \pm 5.95$ & $1.01 \pm 0.095$\\ 
 \hline
 204 & 814 & $-1$ & $401.41 \pm 24.48$ & $-0.22 \pm 0.0029$ & $46.33 \pm 5.95$ & $203.97 \pm 0.29$\\[1ex] 
 \hline
\end{tabular}
 \caption{Swapping algorithm statistics where population 1 comprises $10\%$ and $20\%$ of the network.}
 \label{tbl:stats}
\end{table}

\subsection{The relationship between drive and sortedness with respect to network synchronisation and activation across many initial seeds of the sorting algorithm}

In section \ref{subsec:single_network}, we characterised the behaviours displayed by the networks defined by  the population sets $\mathcal{P}_1$ and $\mathcal{P}_2$ for $G \in [0.3, 0.55]$ (the interval over which cells in population 1 are intrinsically active, whilst those in population 2 are not). 
We found that for strong coupling ($g_{coup}=10$), the threshold for activation and synchronisation of the full network is strongly dependent on $\mathcal{A}$, such that increasing $\mathcal{A}$ decreases the necessary drive $G$ for transition (see \Sec{subsubsec:strong_coupling}). 
For $g_{coup} \in \{1,2\}$, we found several regimes of activity, as described in \Sec{subsubsec:middle_coupling} and \Sec{subsubsec:low_coupling}. 
Here, we wish to establish if the identified domains of activity persist across general families of networks with similar $\mathcal{A}$ but different membership of the population sets.

To do this, we defined ranges for the extracellular glucose concentration $G \in [0.3,0.55]$ and for the number of network iterations $a\in[0,250]$ ($a\in[0,400]$ when ${N_1}/{N} \approx 0.2$).
We selected $M=2,048$ points in the $(a,G)$ plane over these ranges following a Latin hypercube sampling.
For each realisation $m\in \mathbb{N}_M$, we ran \Alg{alg:orig} for $a_m$ iterations and recorded the modified spatial sortedness value $\mathcal{A}_m$.
For maximum coverage over the range of possible values of $\mathcal{A}$, \Alg{alg:orig} was run in either a forward or a backward fashion (see \Sec{sec:algorithm}).
We did this by selecting the Algorithm direction $d_m\in\{-1,1\}$ randomly and with uniform probability.
Once the algorithm terminated, the dynamics \eqref{eq:Veq}-\eqref{eq:coupling} of the resulting network configuration were simulated using the chosen activation value $G_m$ and the summary statistics as described in \Sec{sec:sum_stats} were evaluated.
These summary statistics were then plotted against the set of $(\mathcal{A}_m,G_m)$ values.
We then repeated this process for different values of $g_\text{coup}$ and proportions of population 1 nodes $N_1 = 102$ (${N_1}/{N} \approx 0.1$) (as in the \Sec{subsec:single_network}) and $N_1 = 204$ (${N_1}/{N} \approx 0.2$).

Evaluation of the level sets led to complicated sets due to the use of different realisations of \Alg{alg:orig} and the use of different initial conditions. Since the complex nature of these level sets was not related to the relationship between sortedness, drive, and network dynamics, and further because it obfuscated results, we opted to remove these portions of the levels sets from Figs.~\ref{fig:hypercube_set_strong}-~\ref{fig:hypercube_set_weak}. As an example for comparison, Fig.~\ref{fig:hypercube_set_weak_raw} includes the full level sets corresponding to Fig.~\ref{fig:hypercube_set_weak}A.

\subsubsection{Increasing $\mathcal{A}$ lowers the required drive $G$ for a transition to globally synchronised bursting for strong coupling and varying population sizes}

\begin{figure}[ht] 
\centering
\includegraphics[width=1.0\linewidth]{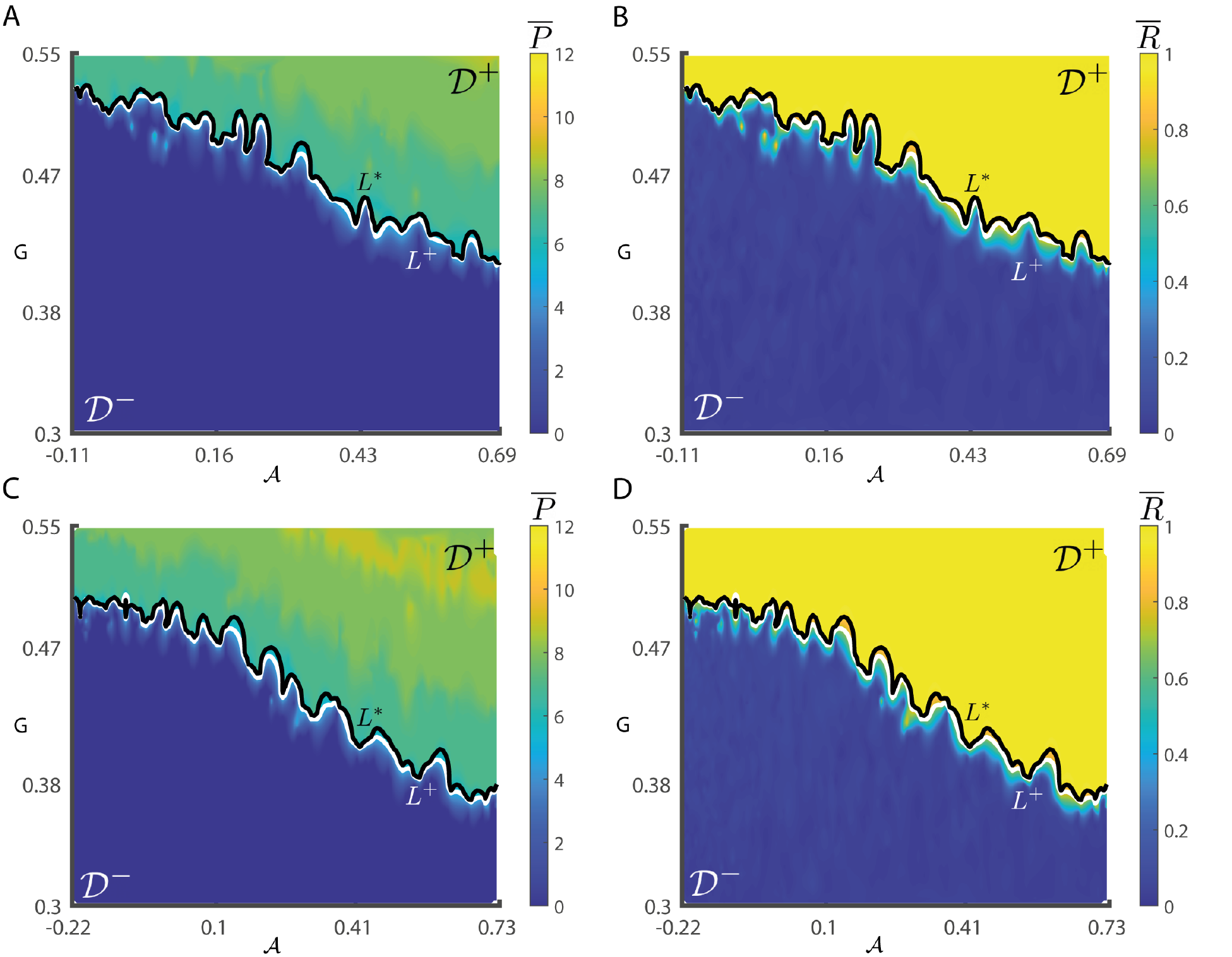}
\caption{\textbf{Network activity with respect to sortedness and drive for strong coupling across realisations of the swapping algorithm.}
\textbf{A)} Plotting $\overline{P}$ on a set of realisations of the swapping algorithm, $(\mathcal{A}_m, G_m)$, shows a decreasing trend in necessary drive with respect to $\mathcal{A}$ for activation. In this case, population 1 was $10\%$ of the network. \textbf{B)} Plotting $\overline{P}$ on a set of realisations of the swapping algorithm, $(\mathcal{A}_m, G_m)$, shows a decreasing trend in necessary drive with respect to $\mathcal{A}$ for synchronisation. In this case, population 1 was $10\%$ of the network. \textbf{C)} As in A, but where population 1 was $20\%$ of the network. \textbf{D)} As in B, but where population 1 was $20\%$ of the network.}
\label{fig:hypercube_set_strong}
\end{figure}

We found that the monotonic decreasing relationship between $\mathcal{A}$ and $G$ discussed in \Sec{subsubsec:strong_coupling} persists when each point $(\mathcal{A}_m,G_m)$ corresponds to a different realisation of the swapping algorithm.
The regimes $\mathcal{D}^-$ and $\mathcal{D}^+$ both exist and can be separated by the same level sets as defined previously: $L^{*} = \{(\mathcal{A}, \ G) \mid \overline{R}(\mathcal{A}, \ G)=0.9\}$, and $L^{+} = \{(\mathcal{A}, \ G) \mid \overline{P}(\mathcal{A}, \ G)=5\}$.
These boundaries show a decreasing trend in $G$ with respect to $\mathcal{A}$ in the transition from global quiescence to globally synchronised oscillations, although due to each point representing a different realisation of the \Alg{alg:orig} (and a distinct set of initial conditions), the separatrix is now non-monotonic.
Fig.~\ref{fig:hypercube_set_strong} shows $\overline{P}$  and $\overline{R}$ when population 1 nodes account for $10\%$ (Fig.~\ref{fig:hypercube_set_strong}A,B) of the network and for $20\%$ (Fig.~\ref{fig:hypercube_set_strong}C,D) of the network. 
We found that increasing the proportion of population 1 nodes did not change the nature of the relationship between $\mathcal{A}$ and $G$, however, the threshold for activation $G$ was decreased over all values of $\mathcal{A}$. 
This decrease in threshold is expected as the number of intrinsically active nodes (and hence `intrinsic' network excitability) in the network was doubled.

\subsubsection{The regime of inter-population resonance persists across realisations of the swapping algorithm for middle-strength coupling}

\begin{figure}[ht] 
\centering
\includegraphics[width=1.0\linewidth]{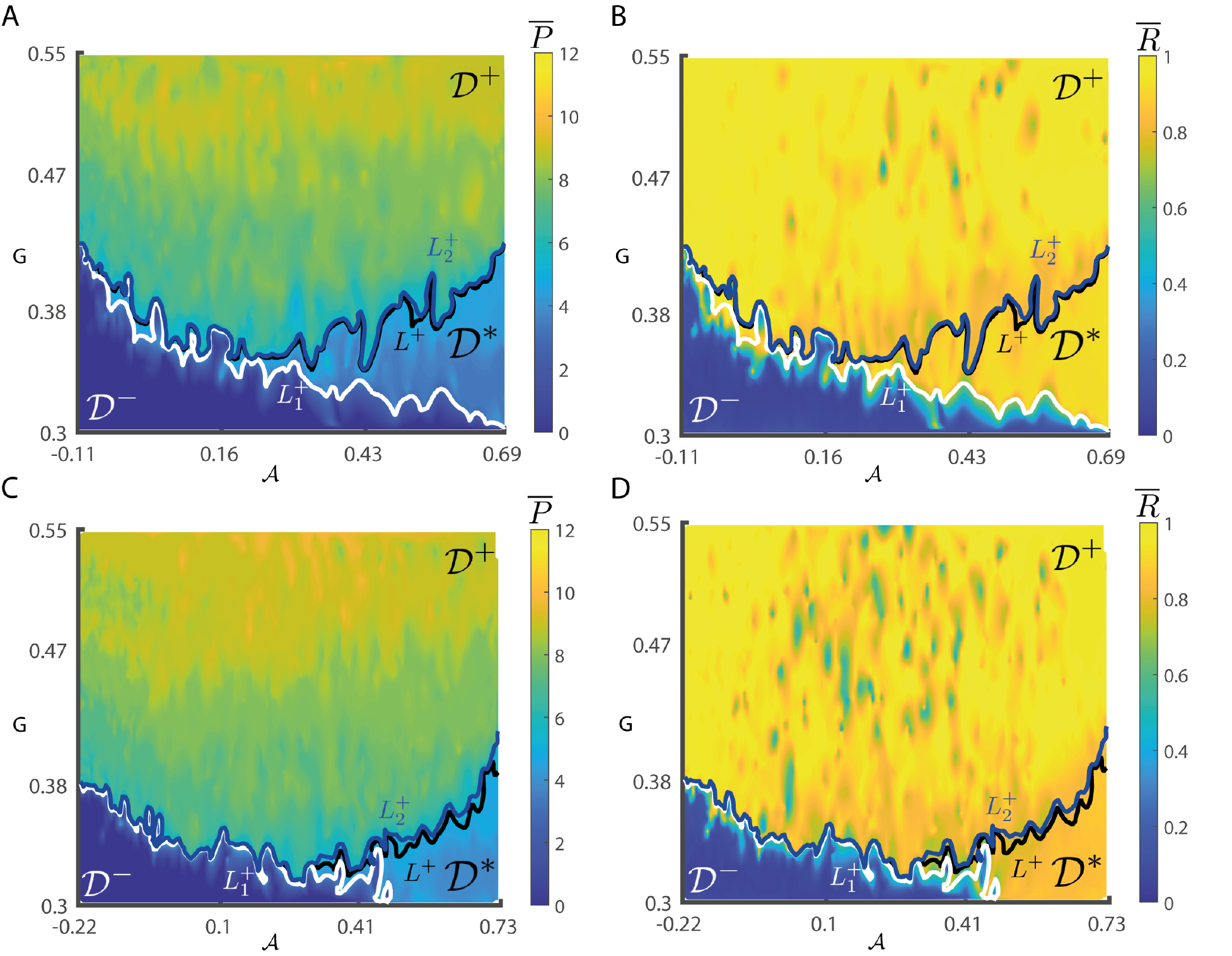}
\caption{\textbf{Network activity with respect to sortedness and drive for middle-strength coupling across realisations of the swapping algorithm.} \textbf{A)} Plotting $\overline{P}$ shows that $\mathcal{D}^*$ persists across realisations of the swapping algorithm. In this case, population 1 was $10\%$ of the network. \textbf{B)} Plotting $\overline{R}$ shows that $\mathcal{D}^*$ persists across realisations of the swapping algorithm. In this case, population 1 was $10\%$ of the network. \textbf{C)} As in A, but where population 1 was $20\%$ of the network. \textbf{D)} As in B, but with where population 1 was $20\%$ of the network.}
\label{fig:hypercube_set_middle}
\end{figure}

For intermediate-strength coupling ($g_{coup}=2$), we found that the regimes discussed in \Sec{subsubsec:middle_coupling} still exist when each point $(\mathcal{A}_m,G_m)$ corresponds to a different realisation of the swapping algorithm.
In particular, we found the existence of the regions $\mathcal{D}^-$, $\mathcal{D}^+$, and $\mathcal{D}^*$, which can be separated by the level sets $L_1^+$ and $L_2^+$, where $L_{k}^{+} =\{(\mathcal{A},G) \mid \overline{P}_{k}(\mathcal{A},G)=5\}$ for $k \in \{1,2\}$, can be used to separate the three regimes. 
Moreover, we observed some network simulations which exhibited lowered $\overline{R}$  within $\mathcal{D}^+$, which we conjecture is the result of multi-stability (i.e., different asymptotic dynamics for different initial conditions), as in Fig.~\ref{fig:network_set_middle_yi}. 
Figure~\ref{fig:hypercube_set_middle} shows $\overline{P}$ and $\overline{R}$ in the case when population 1 nodes comprise $10\%$ (Fig.~\ref{fig:hypercube_set_middle}A,B) and $20\%$ (Fig.~\ref{fig:hypercube_set_middle}C,D) of the network.
As in the case for strong coupling, each regime is shifted downward, with respect to $G$, when the proportion on intrinsically active nodes is increased to $20\%$.
In fact, we found that for high degrees of sortedness, the inter-population resonance regime begins at the lowest value of $G$ that we considered ($G=0.3$). 
This shows that for middle-strength coupling, high sortedness, and where $20\%$ of the network are nodes from population 1, activation of the network occurs for values of $G$ very near where the threshold ($G\approx 0.25)$ at which isolated population 1 nodes become active.

\subsubsection{Non-monotonicity with respect to synchronisation persists for weak coupling across realisations of the swapping algorithm and for differing population 1 sizes}

\begin{figure}[ht] 
\centering
\includegraphics[width=1.0\linewidth]{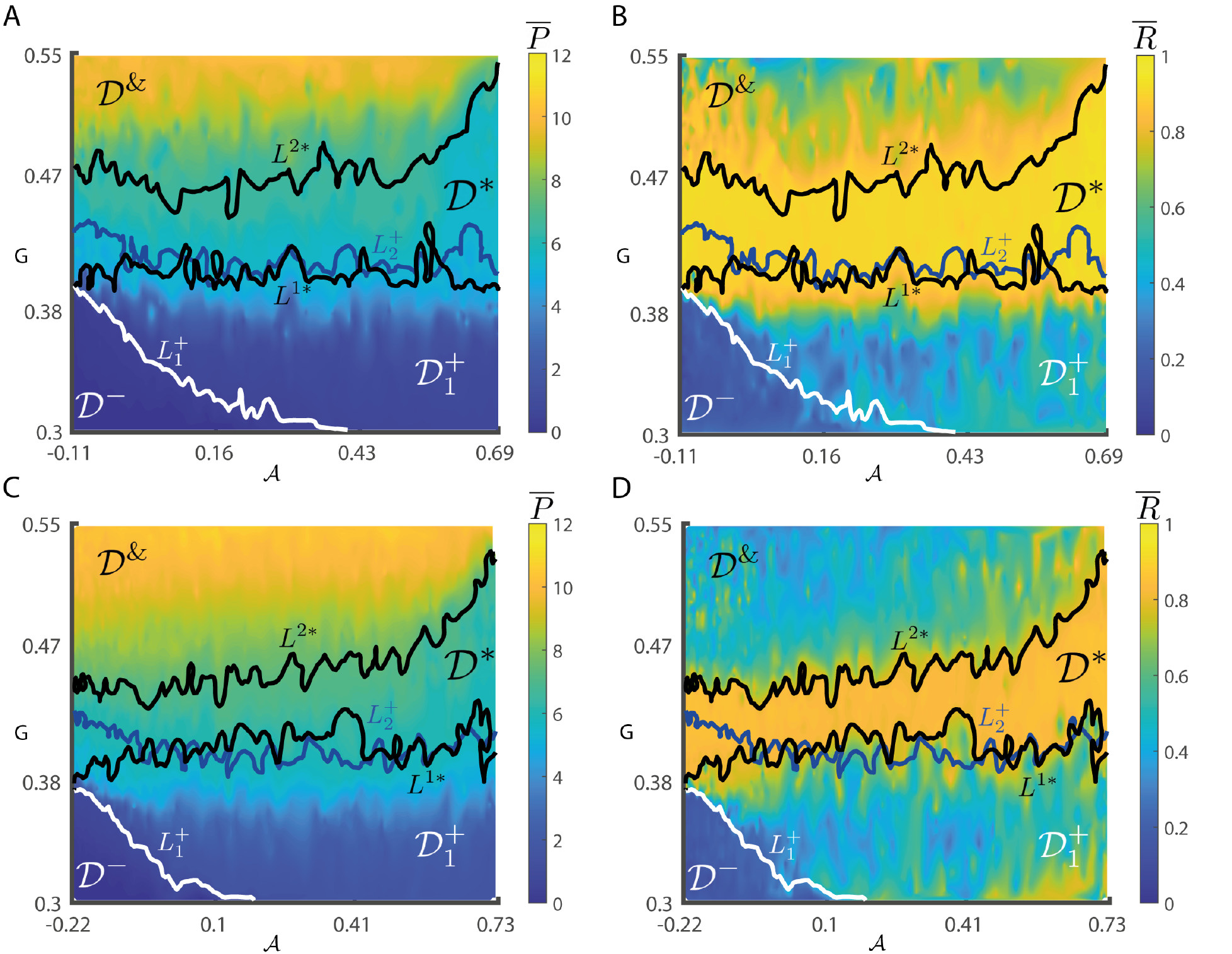}
\caption{\textbf{Network activity with respect to sortedness and drive for weak coupling across realisations of the swapping algorithm.} \textbf{A)} Plotting $\overline{P}$ shows that the regimes $\mathcal{D}^*$ and $\mathcal{D}^\&$ persist across realisations of the swapping algorithm. In this case, population 1 was $10\%$ of the network. \textbf{B)} Plotting $\overline{R}$ shows that the regimes $\mathcal{D}^*$ and $\mathcal{D}^\&$ persist across realisations of the swapping algorithm. In this case, population 1 was $10\%$ of the network. \textbf{C)} As in A, but where population 1 was $20\%$ of the network.\textbf{D)} As in B, but where population 1 was $20\%$ of the network.}
\label{fig:hypercube_set_weak}
\end{figure}

Finally, we considered weak coupling ($g_{coup}=1$) for $(\mathcal{A}_m, G_m)$ using $M$ realisations of the swapping algorithm.
Figure~\ref{fig:hypercube_set_weak} shows $\overline{P}$ and $\overline{R}$ when population 1 nodes account for $10\%$ (Fig.~\ref{fig:hypercube_set_weak}A,B) and $20\%$ (Fig.~\ref{fig:hypercube_set_weak}C,D) of the network. 
We found that non-monotonicity of the boundary to synchronised activity with respect to increasing $G$ was persistent for this weak coupling case.
The upper boundary of the inter-population resonance regime ($\mathcal{D}^*$), given by the set $L^{2*}$, shows an increasing trend with respect to $\mathcal{A}$ both when population 1 node comprise $10\%$ (Fig.~\ref{fig:hypercube_set_weak}B) and $20\%$ (Fig.~\ref{fig:hypercube_set_weak}D) of the network. 
Moreover, we again found that the activation threshold for population 1 nodes with respect to $G$ decreases as $\mathcal{A}$ increases, which is captured by $L_1^+$, where $L_{k}^{+} =\{(\mathcal{A},G) \mid \overline{P}_{k}(\mathcal{A},G)=5\}$ for $k \in \{1,2\}$. 
Interestingly, we found that the activation of population 2 nodes (reflected by $L_2^+$) with respect to $G$ shows a decreasing trend as $\mathcal{A}$ increases, but only for very low values of $\mathcal{A}$. 
We conjecture that this relationship was not observed in \Sec{subsubsec:low_coupling} because only positive values of $\mathcal{A}$ (resulting from the forward algorithm) were considered there, whereas here we also include realisations of the backward algorithm (leading to negative values of $\mathcal{A}$ being considered).
Here, we found that the bounds of $\mathcal{D}^*$, those being $L^{1*}$ and $L^{2*}$, needed to be modified depending on the proportion of population 1 nodes in the network.
In particular, when only $10\%$ of the network nodes were  from population 1, we defined the level set $L^{*} = \{(\mathcal{A}, \ G) \mid \overline{R}(\mathcal{A}, \ G)=0.9\}$ as in \Sec{subsubsec:low_coupling} which subsequently led to the definition of two curves: the lower bound $L^{1*}$ and the upper bound $L^{2*}$ (Fig.~\ref{fig:hypercube_set_weak}B).
However, when the proportion of population 1 nodes was increased to $20\%$, we instead defined $L^{*} = \{(\mathcal{A}, \ G) \mid \overline{R}(\mathcal{A}, \ G)=0.8\}$ (Fig.~\ref{fig:hypercube_set_weak}D). The thresholds we chose were dependent on the number of population 2 nodes. This is because we sought to define level sets that bounded the 2:1 resonance region. In that region, nodes are synchronised within, but not between, populations. Therefore, $\overline{R}$ is approximately equal to the fraction of nodes in the larger population (i.e., population 2).

\subsection{Additional figures referenced in the manuscript}
\begin{figure}[ht] 
\centering
\includegraphics[width=0.9\linewidth]{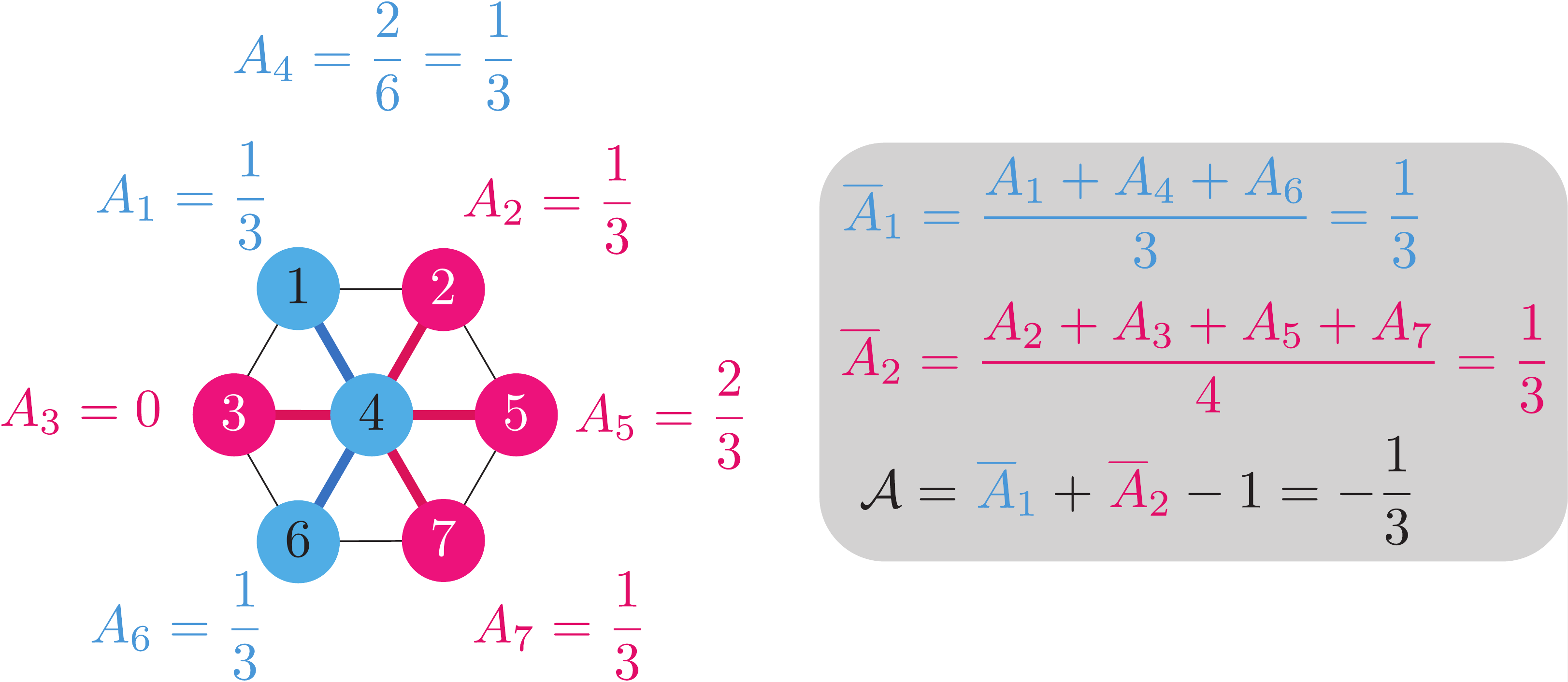}
\caption{\textbf{Illustrative example of network sortedness metric.} The sortedness metrics are computed for the example network comprising $N_1 = 3$ population 1 nodes (blue) and $N_2 = 4$ population 2 nodes (pink) with population sets $P_1 = \{1,4,6\}$ and $P_2 = \{2,3,5,7\}$. The node sortedness values, $A_i$, $i=1,\dots,7$, take the indicated values. For ease of viewing one example calculation, the edges of node 4 are highlighted in the colour corresponding to the population of each of its neighbouring nodes.
The grey box shows the population sortedness evaluations, $\overline{A}_k$, $k\in\{1,2\}$, computed using \eqref{eq:ave_node_assort} and overall network sortedness, $\mathcal{A}$, computed using \eqref{eq:net_assort}.}
\label{fig:assortativity}
\end{figure}

\begin{figure}[t] 
\centering
\includegraphics[width=0.5\linewidth]{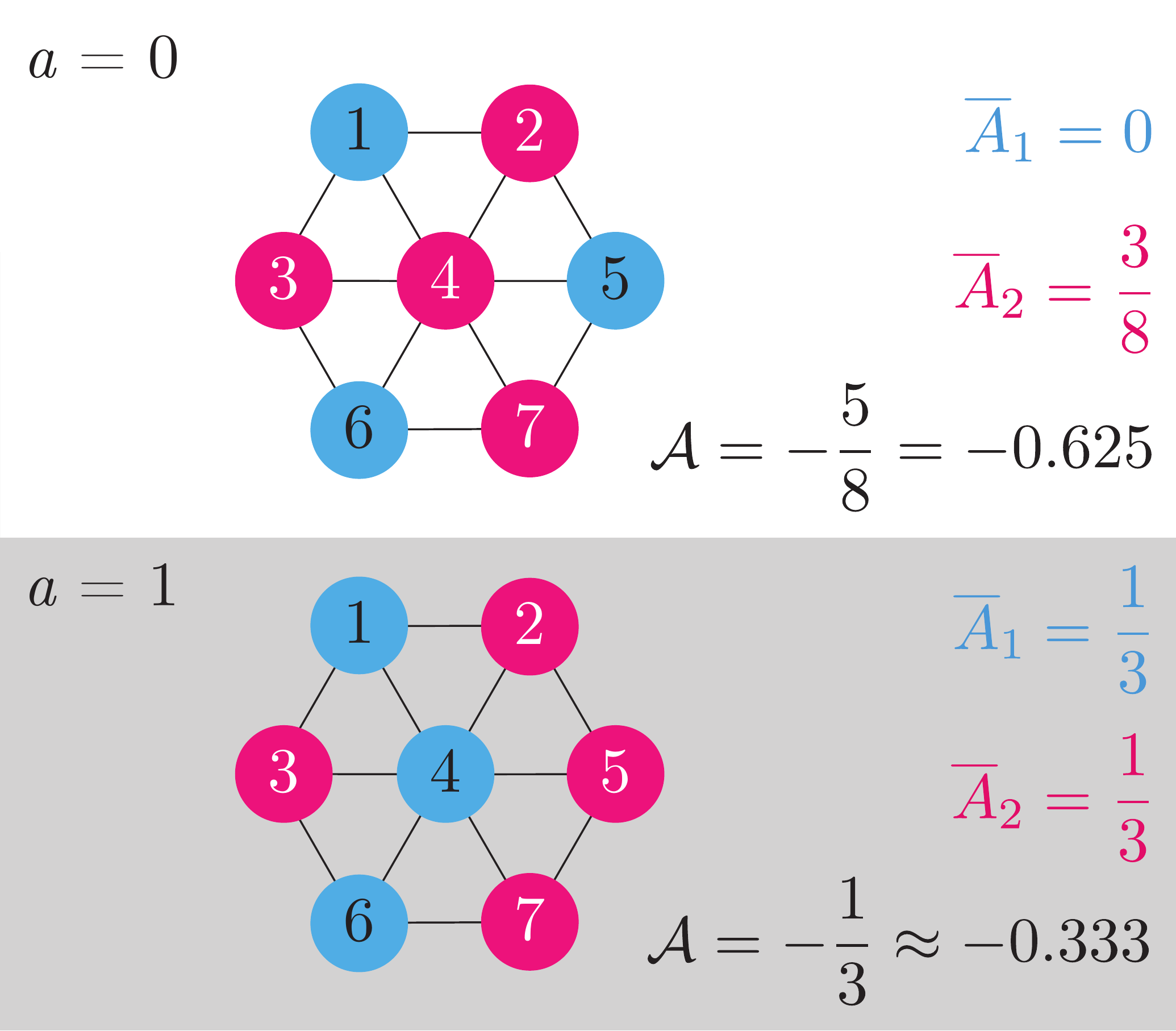}
\caption{\textbf{Example of one iteration of the network sorting algorithm in the forward direction} The initial network with $a=0$ is in the maximally unsorted state so that no two nodes from population 1 (blue) are coupled to one another. Here, the population sets are $P_1 = \{1,5,6\}$ and $P_2 = \{2,3,4,7\}$ and network sortedness is equal to $-5/8$. The algorithm attempts to move node 4 to population 1 and node 5 to population 2 (pink). In the trial configuration shown in the grey box, the network sortedness is equal to $-1/3 > -5/8$ and so the swap is accepted. Thus, the population sets are updated to $P_1 = \{1,4,6\}$ and $P_2 = \{2,3,5,7\}$ and the iteration counter is increased to $a=1$. If the swap were rejected, another pair of nodes would be selected at random and the computation of network sortedness would be repeated. If no possible swap changes $\mathcal{A}$ in the desired direction, the algorithm would terminate without incrementing the iteration number, $a$.}
\label{fig:alg_iteration}
\end{figure}

\begin{figure}[ht]
\centering
\includegraphics[width=1.0\linewidth]{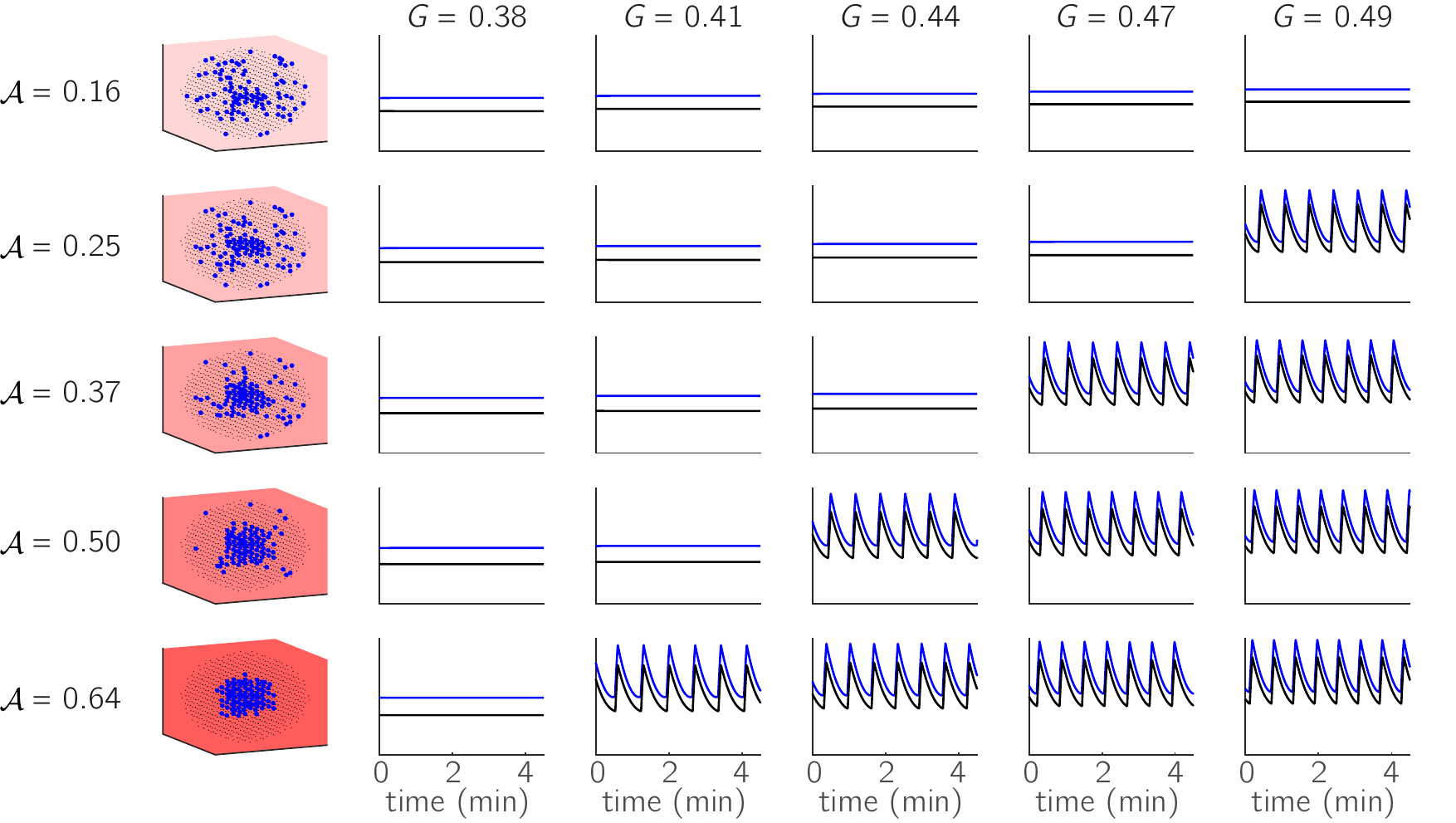}
\caption{\textbf{Phase transitions with respect to spatial sortedness.} The mean Ca\textsuperscript{2+} dynamics of population 1 (blue) and population 2 (black) are shown as $\mathcal{A}$ increases. For higher $G$, activation occurs at lower $\mathcal{A}$. Conversely, increasing $\mathcal{A}$ allows weaker $G$ activate the system.}
\label{fig:phase_transition_example}
\end{figure}

\begin{figure}[ht] 
\centering
\includegraphics[width=1.0\linewidth]{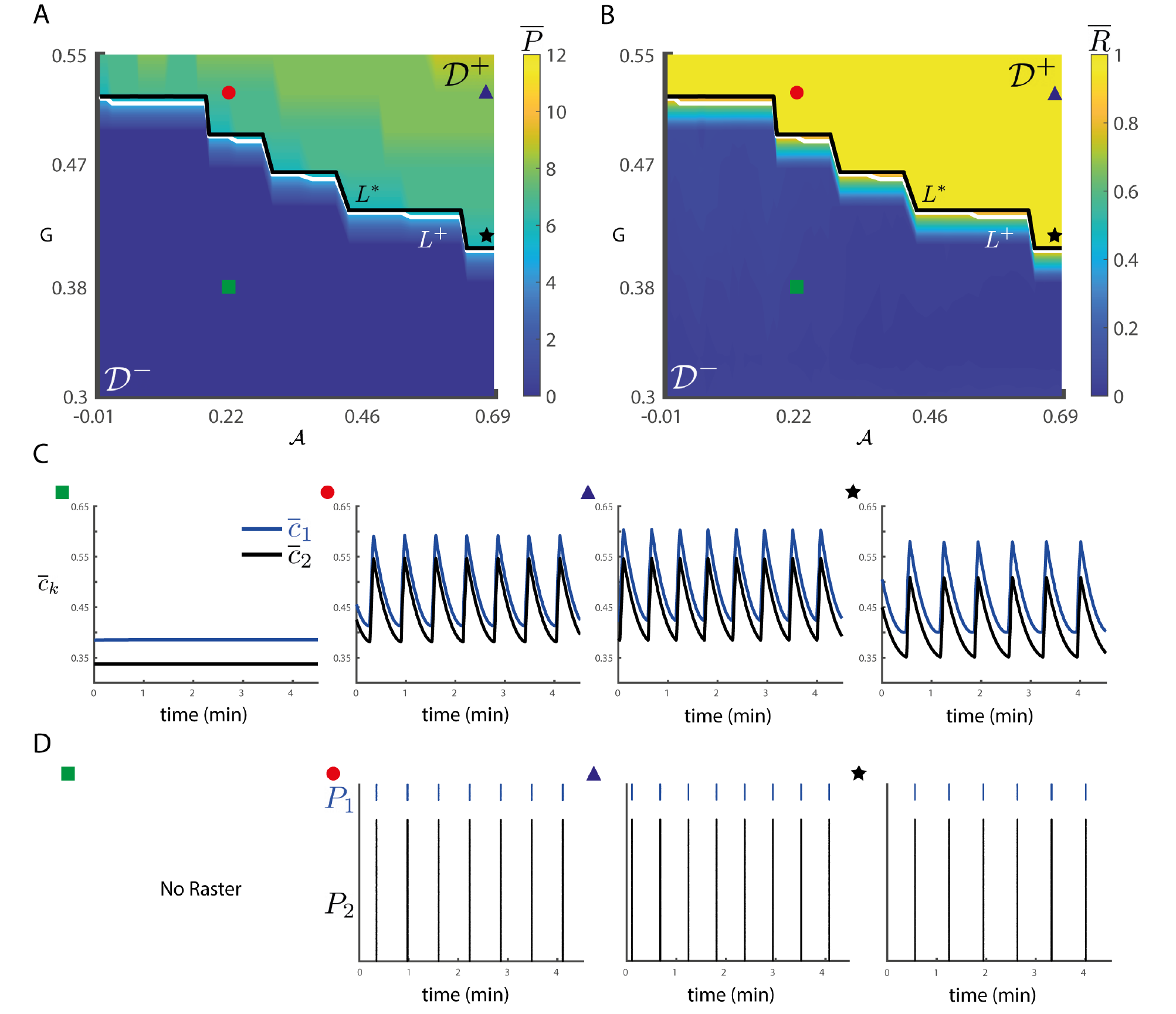}
\caption{\textbf{Network activity with respect to sortedness and drive for strong coupling.} \textbf{A)} Plotting $\overline{P}$ averaged over three sets of initial conditions shows that for increasing $\mathcal{A}$, lower drive $G$ is required to activate the network. \textbf{B)} Plotting $\overline{R}$ averaged over three sets of initial conditions shows that for increasing $\mathcal{A}$, lower drive $G$ is required to synchronise the network. \textbf{C)} Average Ca\textsuperscript{2+} across population 1 nodes ($\overline{c}_1$) and population 2 nodes ($\overline{c}_2$) for ($\mathcal{A}$, $G$) pairs illustrates a strong global signal in $\mathcal{D}^+$. \textbf{D}) Raster plots showing the strong synchronisation within $\mathcal{D}^+$ for strong coupling. The raster plot is ordered such that nodes whose indices are in $P_1$, i.e. population 1 nodes, are shown in blue at top of the plot, whilst nodes whose indices are in $P_2$ are shown in black at the bottom of the plot.}
\label{fig:network_set_strong}
\end{figure}

\begin{figure}[ht] 
\centering
\includegraphics[width=0.95\linewidth]{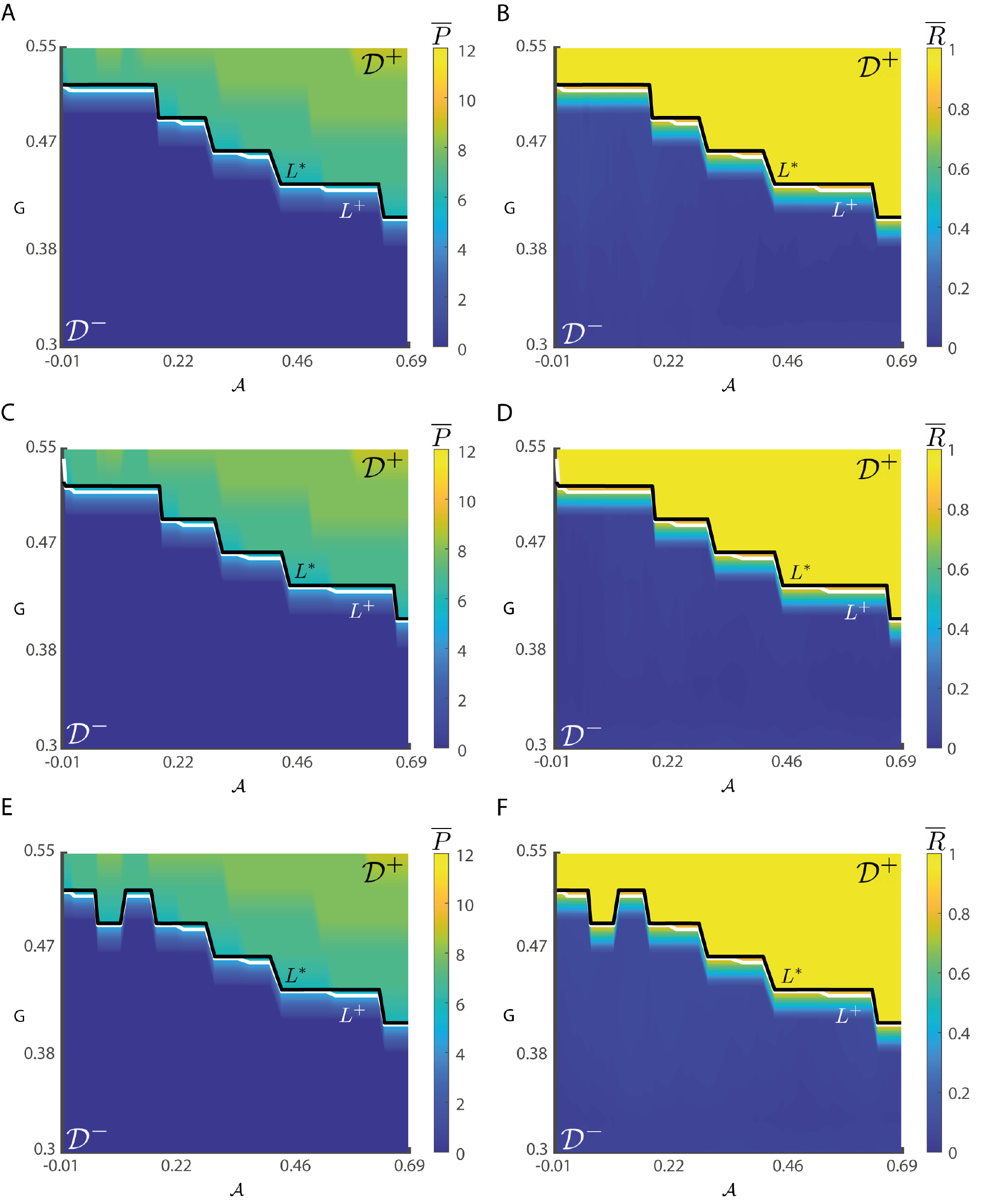}
\caption{\textbf{Network activity with respect to sortedness and drive for strong coupling (three initial conditions).} \textbf{A)} Plotting $\overline{P}$ shows that for increasing $\mathcal{A}$, lower drive $G$ is required to activate the network (parameter set $Y_1(0)$). \textbf{B)} Plotting $\overline{R}$ shows that for increasing $\mathcal{A}$, lower drive $G$ is required to synchronise the network (parameter set $Y_1(0)$). \textbf{C)} $\overline{P}$ for $Y_2(0)$. \textbf{D)} $\overline{R}$ for $Y_2(0)$. \textbf{E)} $\overline{P}$ for $Y_3(0)$. \textbf{F)} $\overline{R}$ for $Y_3(0)$.}
\label{fig:network_set_strong_yi}
\end{figure}

\begin{figure}[ht] 
\centering
\includegraphics[width=1.0\linewidth]{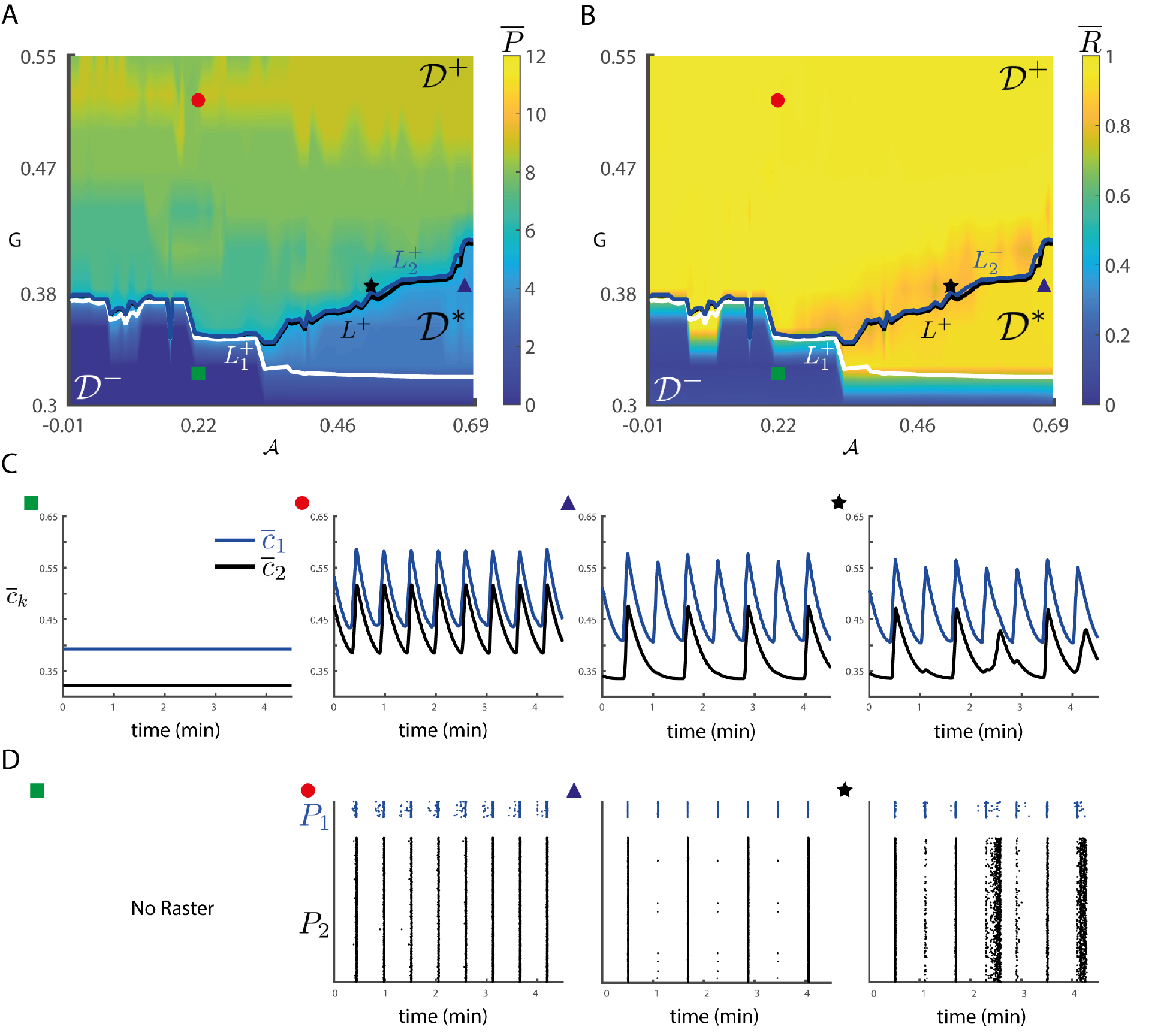}
\caption{\textbf{Network activity with respect to sortedness and drive for middle-strength coupling.} \textbf{A)} Plotting $\overline{P}$ averaged over three sets of initial conditions shows a third regime $\mathcal{D}^*$ bounded by $L_1^+$ and $L_2^+$. \textbf{B)} Plotting $\overline{R}$ averaged over three sets of initial conditions shows that for increasing $\mathcal{A}$, lower drive $G$ is required to synchronise the network. \textbf{C)} Average Ca\textsuperscript{2+} across population 1 nodes ($\overline{c}_1$) and population 2 nodes ($\overline{c}_2$) for ($\mathcal{A}$, $G$) pairs illustrates a strong global signal in $\mathcal{D}^+$ and that population 2 nodes are active at half the frequency of population 1 nodes, on average, in $\mathcal{D}^*$. \textbf{D}) Raster plots showing the strong synchronisation within $\mathcal{D}^+$, 2:1 frequency resonance in $\mathcal{D}^*$, and intermediate activity with lowered synchronisation in a band separating the two regimes. The raster plot is ordered such that nodes whose indices are in $P_1$, i.e. population 1 nodes, are shown in blue at top of the plot, whilst nodes whose indices are in $P_2$ are shown in black at the bottom.}
\label{fig:network_set_middle}
\end{figure}

\begin{figure}[ht] 
\centering
\includegraphics[width=0.7\linewidth]{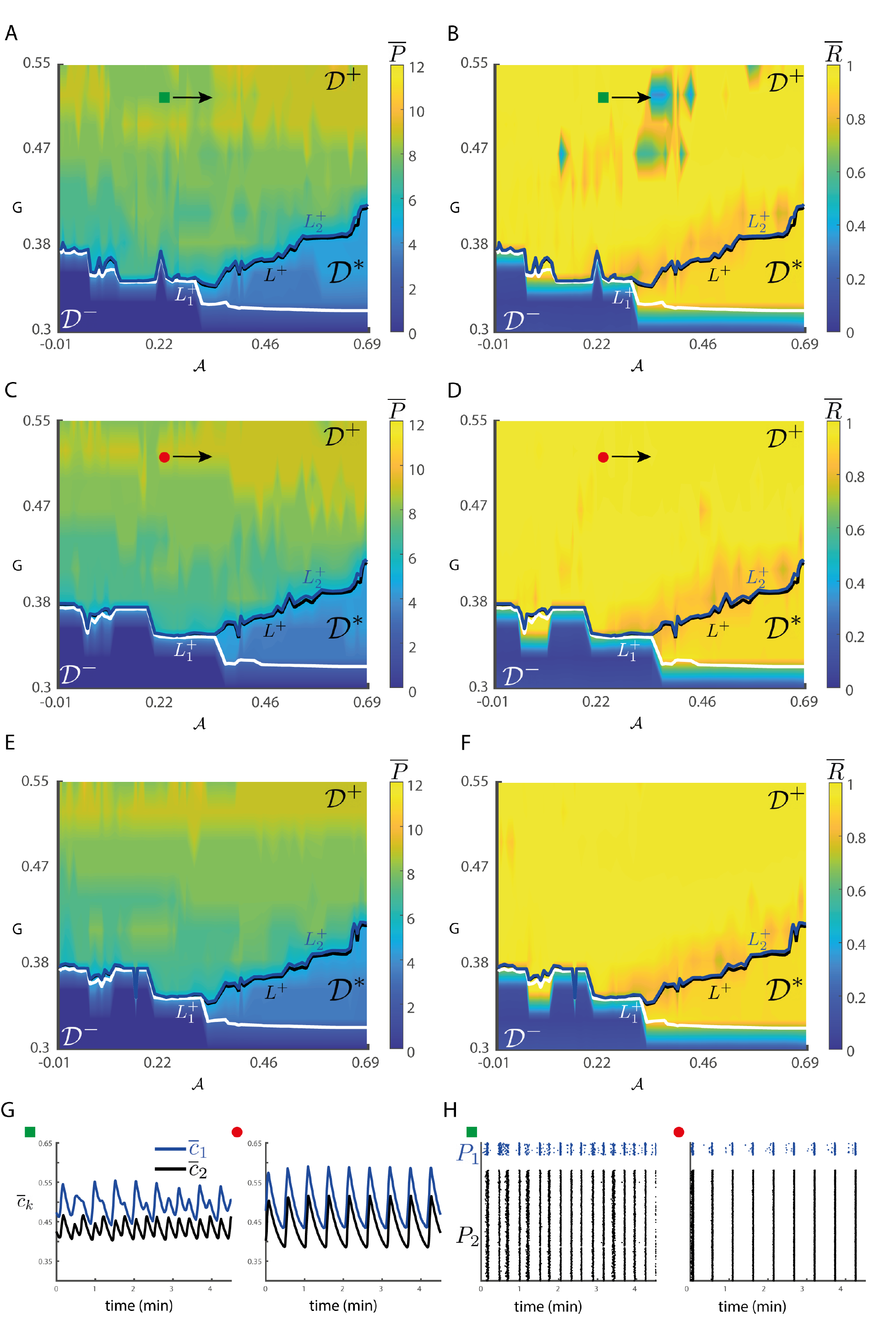}
\caption{\textbf{Network activity with respect to sortedness and drive for middle-strength coupling (three initial conditions).} \textbf{A)} $\overline{P}$ for $Y_1(0)$. \textbf{B)} $\overline{P}$ for $Y_1(0)$. \textbf{C)} $\overline{P}$ for $Y_2(0)$. \textbf{D)} $\overline{R}$ for $Y_2(0)$. \textbf{E)} $\overline{P}$ for $Y_3(0)$. \textbf{F)} $\overline{R}$ for $Y_3(0)$. \textbf{G)} Mean Ca\textsuperscript{2+} for $P_1$ ($\overline{c}_1$) and $P_2$ ($\overline{c}_2$) showing multi-stability in the $\mathcal{D}^+$ regime. \textbf{H)} Raster plots of the population 1 (blue) and population 2 (black) nodes showing multistability in the $\mathcal{D}^+$ regime. The raster plot is ordered such that nodes whose indices are in $P_1$, i.e. population 1 nodes, are shown at the top of the plot.}
\label{fig:network_set_middle_yi}
\end{figure}

\begin{figure}[ht] 
\centering
\includegraphics[width=1.0\linewidth]{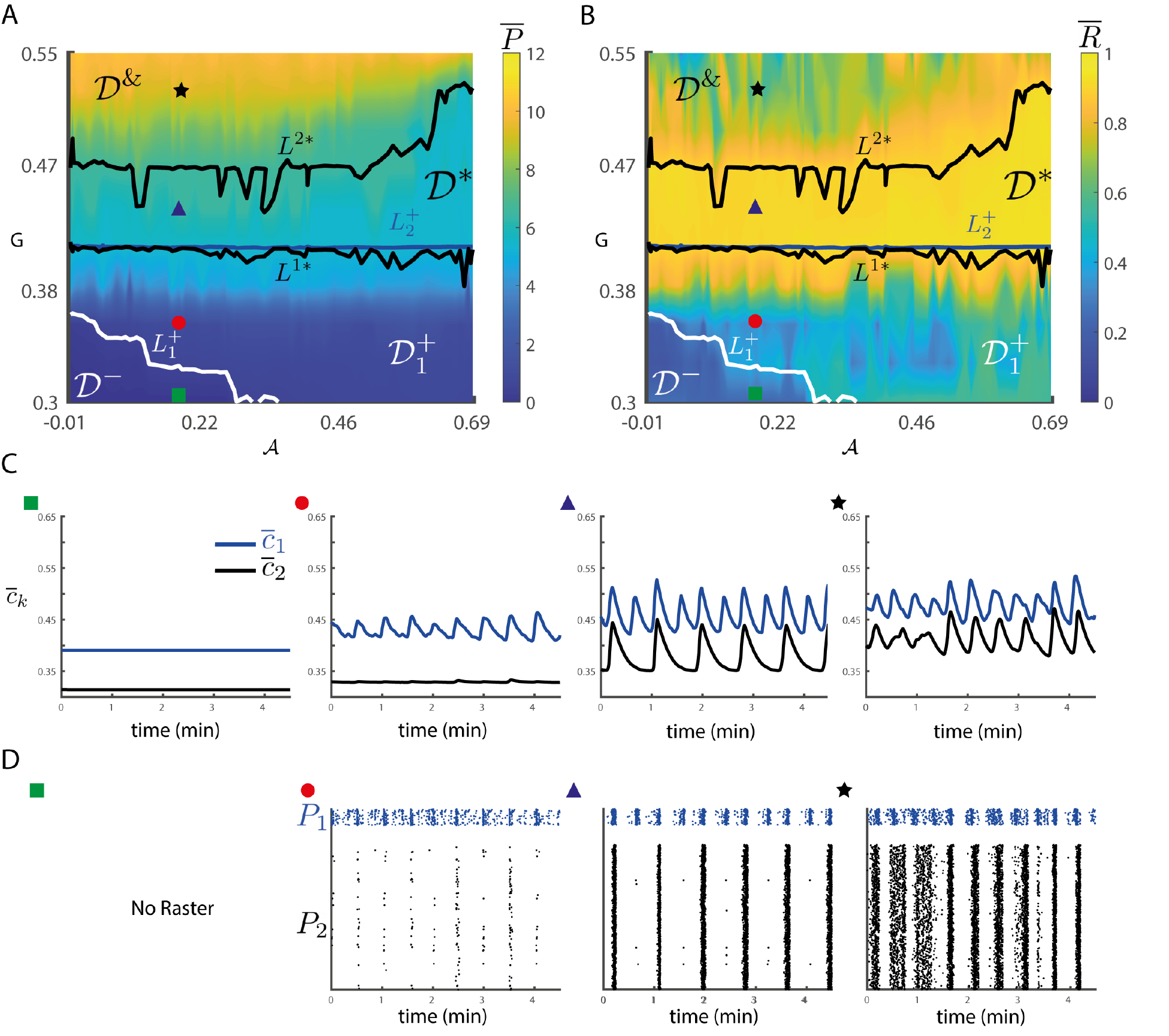}
\caption{\textbf{Network activity with respect to sortedness and drive for weak coupling.} \textbf{A)} Plotting $\overline{P}$ averaged over three sets of initial conditions shows that activation of population 1, but not population 2, is dependent on sortedness. \textbf{B)} Plotting $\overline{R}$ averaged over three sets of initial conditions shows that synchronisation is non-monotonic with respect to $G$, peaking within a $2:1$ resonance regime $\mathcal{D}^*$. \textbf{C)} Average Ca\textsuperscript{2+} for population 1 nodes ($\overline{c}_1$) and population 2 nodes ($\overline{c}_2$) for ($\mathcal{A}$, $G$) shows that population 1 is active but only generates a weak global signal in $\mathcal{D}_{1}^{+}$. The dynamics exhibit 2:1 resonance within the region $\mathcal{D}^*$, and lowered coordination and an irregular global signal within $\mathcal{D}^\&$. \textbf{D}) Raster plots showing the weak coordination of spiking activity across population 1 in $\mathcal{D}_{1}^{+}$ and weak coordination of spiking activity across the whole network within $\mathcal{D}^\&$. The raster plot is ordered such that nodes whose indices are in $P_1$, i.e. population 1 nodes, are shown in blue at top of the plot, whilst nodes whose indices are in $P_2$ are shown in black at the bottom.}
\label{fig:network_set_weak}
\end{figure}

\begin{figure}[ht] 
\centering
\includegraphics[width=0.95\linewidth]{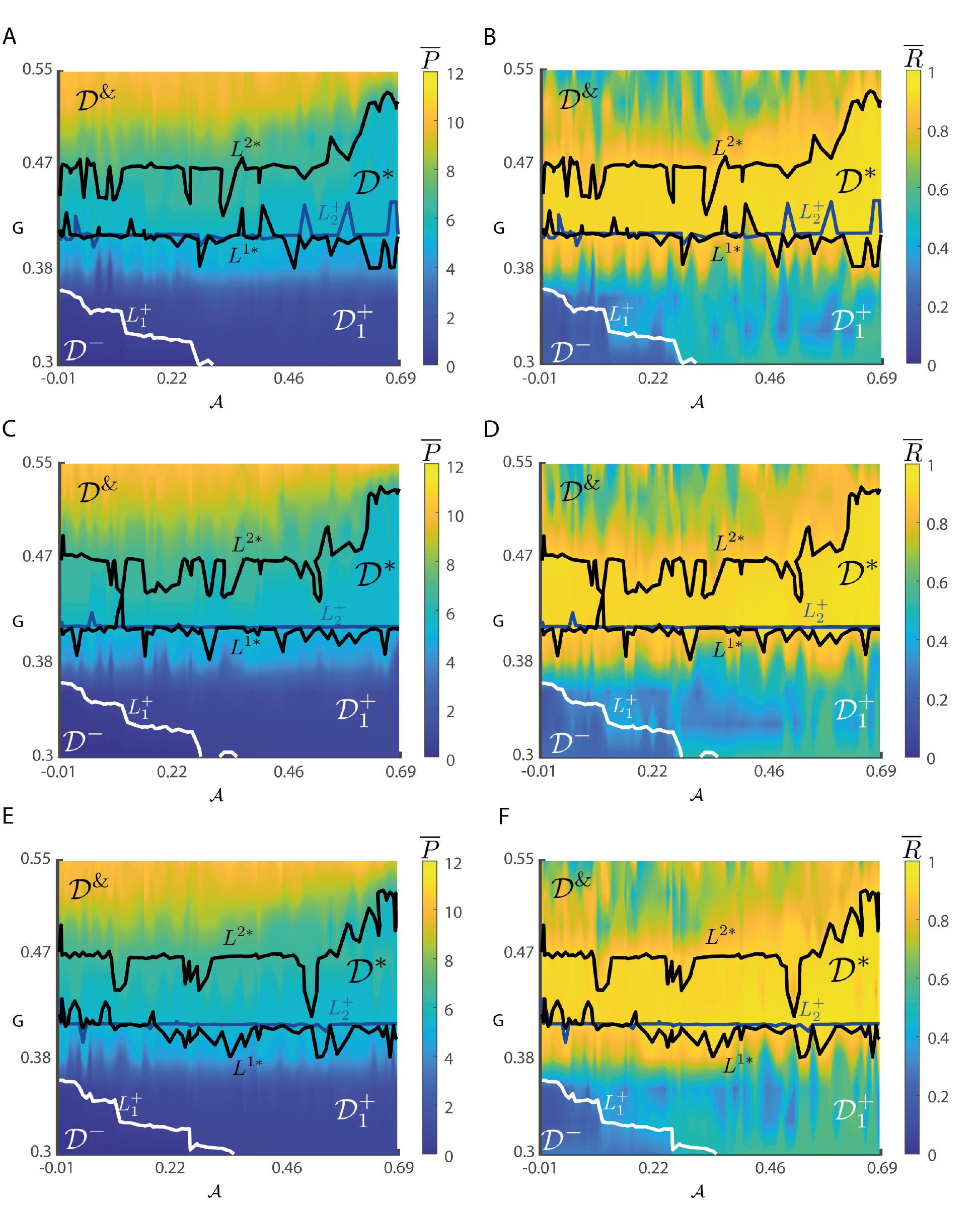}
\caption{\textbf{Network activity with respect to sortedness and drive for weak coupling (three initial conditions).} \textbf{A)} $\overline{P}$ for $Y_1(0)$. \textbf{B)} $\overline{P}$ for $Y_1(0)$. \textbf{C)} $\overline{P}$ for $Y_2(0)$. \textbf{D)} $\overline{R}$ for $Y_2(0)$. \textbf{E)} $\overline{P}$ for $Y_3(0)$. \textbf{F)} $\overline{R}$ for $Y_3(0)$.}
\label{fig:network_set_weak_yi}
\end{figure}

\begin{figure}[ht] 
\centering
\includegraphics[width=1.0\linewidth]{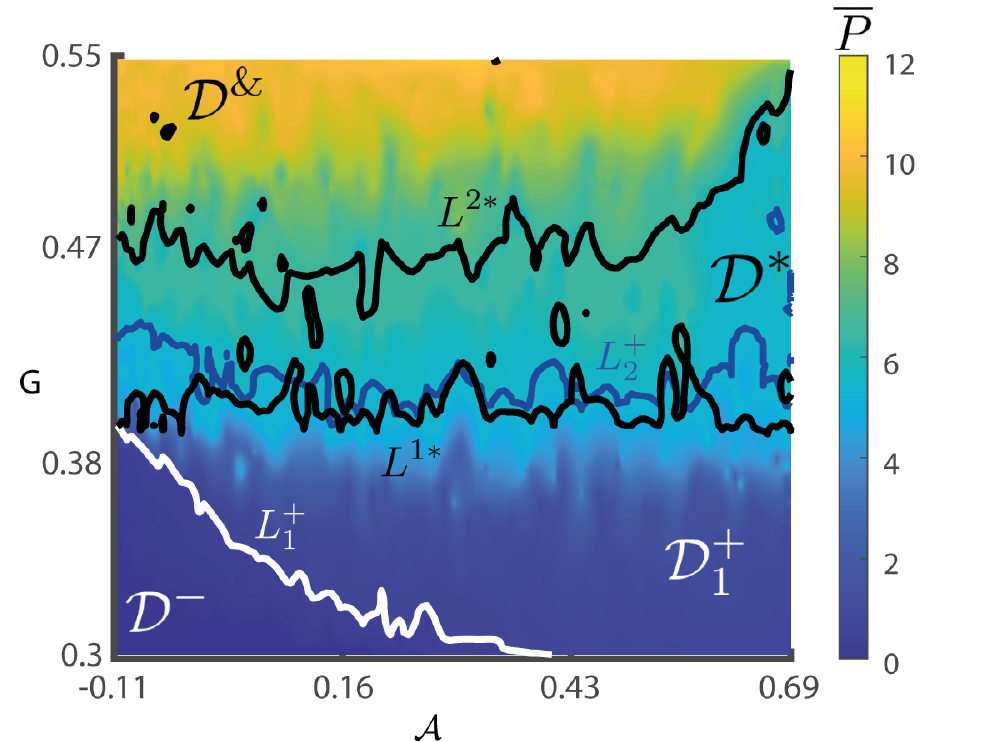}
\caption{\textbf{Raw version of Fig.~\ref{fig:hypercube_set_weak}A.} This figure shows the complete level sets. We only kept the portions of the level sets $L_1^+$, $L_2^+$, $L^{1*}$, and $L^{2*}$ that had analogues in Fig.~\ref{fig:network_set_weak}A.}
\label{fig:hypercube_set_weak_raw}
\end{figure}

\end{document}